\documentclass[aps,twocolumn,superscriptaddress]
{revtex4-1}
\usepackage{xspace,amsmath,amsfonts,amssymb,amsbsy}
\usepackage{ntheorem}
\usepackage{graphicx,epstopdf}
\usepackage{bbm}
\usepackage{footnote}
\usepackage[usenames,dvipsnames]{xcolor}
\usepackage[breaklinks=true]{hyperref}
\hypersetup{
colorlinks   = true, %Colours links instead of ugly boxes
urlcolor     = blue, %Colour for external hyperlinks
linkcolor    = blue, %Colour of internal links
citecolor    = red %Colour of citations
}
\usepackage{soul}
\usepackage{ulem}
\begin{document}


\title{Control  and amplification of Bloch oscillations via photon-mediated interactions}
% Cavity mediated interaction as a toolbox to control the Bloch oscillation
% Many-body control of Bloch oscillation in an optical cavity 
\author{Haoqing Zhang}
\affiliation{JILA, NIST and Department of Physics, University of Colorado, Boulder, Colorado 80309, USA}
\affiliation{Center for Theory of Quantum Matter, University of Colorado, Boulder, Colorado 80309, USA}
\author{Anjun Chu}
\affiliation{JILA, NIST and Department of Physics, University of Colorado, Boulder, Colorado 80309, USA}
\affiliation{Center for Theory of Quantum Matter, University of Colorado, Boulder, Colorado 80309, USA}
\author{Chengyi Luo}
\affiliation{JILA, NIST and Department of Physics, University of Colorado, Boulder, Colorado 80309, USA}
\author{James K. Thompson}
\affiliation{JILA, NIST and Department of Physics, University of Colorado, Boulder, Colorado 80309, USA}
\author{Ana Maria Rey}
\affiliation{JILA, NIST and Department of Physics, University of Colorado, Boulder, Colorado 80309, USA}
\affiliation{Center for Theory of Quantum Matter, University of Colorado, Boulder, Colorado 80309, USA}

\date{\today}

\begin{abstract}
We propose a scheme to control and enhance atomic Bloch oscillations via photon-mediated interactions in an optical lattice supported by a standing-wave cavity with incommensurate lattice and cavity wavelengths. 
% The  scheme  does not require to prepare   a quantum degenerate gas but instead uses   position-dependent atom-light couplings to prepare atoms at specific lattice sites to initialize the dynamics. 
% \textcolor{blue}{\sout{Our scheme uses position-dependent atom-light couplings to spatially  prepare, from a  thermal gas, to an array of  atoms at specific lattice sites.}}
Our scheme uses position-dependent atom-light couplings in an optical cavity to spatially prepare an array of atoms at targeted lattice sites starting from a thermal gas.
 On this initial state we take advantage of  dispersive position-dependent atom-cavity couplings to perform  non-destructive measurements of single-particle Bloch oscillations,  and to generate long-range interactions self-tuned by atomic motion. The latter leads to the generation of  dynamical phase transitions in the deep lattice regime and the amplification of Bloch oscillations in the shallow lattice regime. Our work introduces new possibilities accessible in state-of-the-art cavity QED  experiments  for the  exploration of many-body dynamics in self-tunable potentials.
%thanks to  the interplay between atomic motion, gravity and cavity-mediated interactions.
\end{abstract}
\maketitle

\section{Introduction} 
Bloch oscillations (BO)~\cite{bloch1929} are center-of-mass oscillations or coherent breathing experienced by independent particles in a periodic lattice potential in the presence of a constant force (e.g. gravity).
%Conceptually BO are a beautiful demonstration    of the profound  consequences  of periodic  confinement   on the single particle dynamics.   
Although it has been hard to directly  control BO in conventional electron systems, they have been observed in tailored semiconductor systems~\cite{bo0} as well as ultracold atom systems trapped in optical lattices~\cite{dahan1996bloch,anderson1998macroscopic}. Nevertheless, for the latter, the lattice potential is by implementation  rigid and therefore not a good test bed example of the underlying physics in real materials where the phonons of the crystal dynamically interact with the electron motion. Furthermore, 
inter-atomic interactions have  always  been  a competing mechanism which  damp the oscillations.

Here we propose a scheme to  control and amplify atomic BO via photon-mediated interactions in a gravity-tilted optical lattice supported by a standing-wave optical cavity with  incommensurate lattice and cavity wavelengths. In our case, photons can actively modify the periodic potential experienced by the atoms and therefore resemble the role of phonons in a real solid state environment. 
Even though  experiments that track  BO in optical cavities have been implemented before using a Bose Einstein Condensate (BEC)~\cite{Hemmerich1,Hemmerich2,peden2009,bo8,venk2013}, here  we propose to use inhomogeneous atom-light couplings to prepare  an array of atoms  on  specific lattice sites and  initialize the dynamics \cite{wu2021site}. This can be achieved via position-dependent dispersive atom-light couplings to map the motion of the  atoms under BO into the frequency shift of the cavity resonance. 
Our protocol not only avoids the  ultracold degenerate initial states required in non-destructive measurements of BO, but also provides flexible self-tunability of the cavity-mediated long-range interactions by the atomic motion.  Moreover, in contrast to prior experiments  where the periodic potential was generated by the probe laser field itself~\cite{Hemmerich1,Hemmerich2,peden2009,bo8,venk2013} or separate probe field for site-independent atom-light coupling~\cite{bo11}, we use an additional lattice potential that traps the atoms and controls the degree of delocalization  of the underlying Wannier-Stark (WS) states~\cite{dpt3}  in our system. In this setting, different  to the well-studied case of contact interactions ~\cite{bo1,bo2,bo3,bo4,bo5,bo6,bo9,bo10},   the  photon-mediated interactions can  modify BO depending on the position of other atoms  in the array. Taking advantage of this feature we show versatile many-body phenomena  can be realized in different parameter regimes of this system: 
% such as dynamical phase transitions (DPT) in the deep lattice regime and the  amplification of Bloch oscillations in the shallow lattice regime.
In the deep lattice region, we find dynamical phase transitions (DPT) related to the Lipkin-Meshkov-Glick (LMG) model~\citep{dpt1,dpt2}, which potentially enables rapid generation of spin-squeezed states~\citep{ma2011quantum,pezze2018quantum,li2022improving} with WS states directly, bypassing the need for Raman transitions~\citep{dpt3}; In the shallow lattice, we find the amplification of Bloch oscillation amplification originating from the pair production~\citep{gross2011atomic,lucke2011twin,periwal2021programmable,finger2023spin}  process from the central to adjacent WS states.
We also discuss feasible implementations in state-of-the-art cavity QED experiments~\citep{panda2022quantum,luo2023cavity}.

%Here, we present a scheme to study ``Bloch oscillation" in a tilted lattice support by the optical cavity. 
%The QND measurements are not only used as a probe but also for the cavity-mediated all to all interaction. The basic idea is to use the delocalized Wannier-Stark (WS) state \cite{gluck2002wannier,dpt3} with an equal energy spacing between the eigenstate. The interference between WS states has the Bloch frequency dependent and forms the coherence breathing mode. We work with the region the cavity field can be adiabatic eliminated and form a dynamical potential follow the atomic motion, which is incommensurate with the titled lattice. We explore different lattice depth and cavity parameters and find that the single particle BO with cavity-mediated long-range interaction provides rich dynamics behaviors, which are absent in the contact interaction. The ability to tune the coupling between the atoms and cavity mode that mediates interactions and introduces disorder under the gravity field opens new possibilities in the quantum many-body simulators.

%%%%%%%%%%%%%%%%%%%%%%%%%%%%%%%%%%%%%%%%%%%%%%%%%%%%%%%%%%%%%%%
\begin{figure}[t]
	\centering
	\includegraphics[width=1\columnwidth]{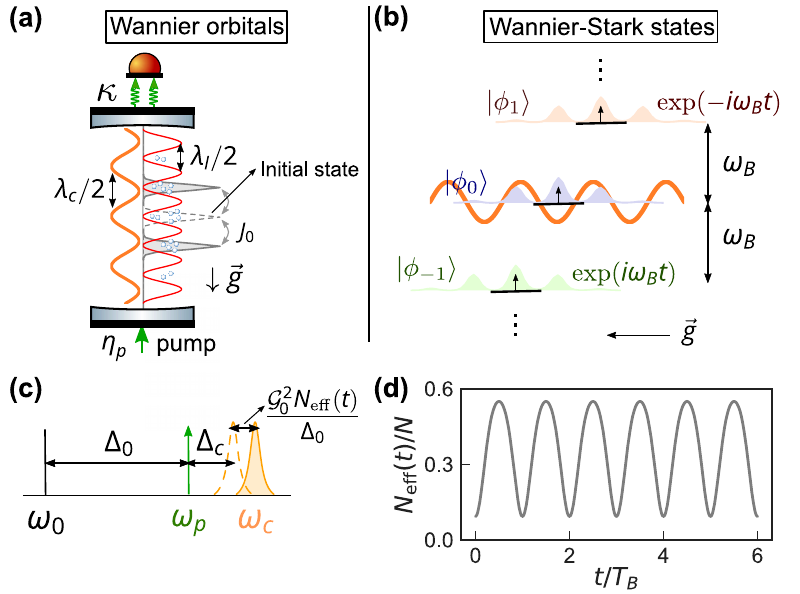}
	\caption{\label{fig:setup} Model system. (a) An ensemble of $N$ atoms are trapped in the lowest band of an optical lattice supported by an optical cavity aligned with gravitational acceleration $\vec{g}$. Considering the atoms are initially localized in a Wannier orbital (grey dashed line), hopping to the nearby sites (grey solid line) can lead to a change of atom-cavity coupling due to incommensurate lattice ($\lambda_l$) and cavity ($\lambda_c$) wavelengths. The cavity has a finite linewidth $\kappa$. (b) The initially localized Wannier orbitals can also be written as a superposition of partially delocalized Wannier-Stark states which accumulate different phases due to gravity. (c) Frequencies of atomic transition ($\omega_0$), external pump ($\omega_p$) and cavity resonance ($\omega_c$). Due to atomic motion, the cavity resonance will be shifted by $\mathcal{G}_0^2N_{\mathrm{eff}}(t)/\Delta_0$, with $N_{\mathrm{eff}}(t)$ defined in Eq.~\eqref{eq:3}. (d) $N_{\mathrm{eff}}(t)$ displays oscillatory behavior reflecting single-particle atomic BO, generated by a sudden quench on lattice depth from $15E_R$ to $8E_R$.}
\end{figure}
%%%%%%%%%%%%%%%%%%%%%%%%%%%%%%%%%%%%%%%%%%%%%%%%%%%%%%%%%%%%%%%

\section{Model} 
We consider an ensemble of $N$ ultracold atoms with mass $M$ trapped in a standing-wave optical cavity along the vertical direction $\hat{z}$ as shown in Fig.~\ref{fig:setup}(a). The atoms are confined in the lowest band of the one-dimensional (1D) optical lattice supported by the cavity, with local gravitational acceleration $\vec{g}$ generating an additional $Mgz$ potential between sites separated by a vertical distance $z$. Here we consider the pure 1D model for simplicity and discuss the modification by the radial modes in \footnote{See Supplemental Material at [URL will be inserted by publisher] for details of effective Hamiltonian derivation, dynamical phase transition, undepleted pump approximation and experimental implementation, includes Ref. \cite{bo9,dpt1,dpt2,dpt3,cox2016spatially}}.  A single internal level $\left|g\right\rangle$ in the atomic ground manifold, is coupled to an atomic excited state $|e\rangle$ with a transition energy $\hbar \omega_0$ via a single cavity mode $\hat{a}$ with frequency $\omega_c$ and wavelength $\lambda_c$. 
The atom-cavity coupling has spatial dependence $\mathcal{G}(z)=\mathcal{G}_0 \sin(k_c z)$,  where $k_c=2\pi/\lambda_c$ and $\mathcal{G}_0$ is proportional to single atom vacuum Rabi splitting.
The cavity mode is coherently driven by an additional laser with frequency $\omega_p$ thus detuning $\Delta_c=\omega_p-\omega_c$ from the bare cavity mode, which generates a net injected field in the cavity with amplitude $\eta_p$. The cavity has a finite linewidth $\kappa$. 

We work in the dispersive regime of the atom-light interaction, where both the cavity mode and the external drive are far detuned from the atomic resonance, i.e., $\Delta_0 \gg \mathcal{G}_0 \sqrt{\left\langle \hat{a}^\dagger \hat{a} \right\rangle }$ with $\Delta_0=\omega_p - \omega_0$. In such limit, we can adiabatically eliminate the excited state and only consider the atomic motion in the ground state, which results in the following second-quantized Hamiltonian,
\begin{equation}
  \hat{H}=\hat H_0 + \int d z \hat{\psi}_g^{\dagger}(z)\frac{\hbar\left|\mathcal{G}(z)\right|^{2}}{\Delta_0} \hat{a}^{\dagger} \hat{a} \hat{\psi}_g(z)+\hat{H}_{\mathrm{cav}}, 
\label{eq:1}
\end{equation}
where $\hat{H}_0=\int d z \hat{\psi}_g^{\dagger}(\hat{p}^{2}/2 M+V_{0} \sin ^{2}\left(k_{l} z\right)+M g z)\hat{\psi}_g$ includes the kinetic energy, lattice potential, and gravitational potential experienced by the atoms. Here, $V_0$ is the lattice depth, $k_l=2\pi/\lambda_l$ is the wavenumber of the lattice beam that sets the atomic recoil energy $E_R=\hbar^2k_l^2/2M$, where $\lambda_l$ is the lattice wavelength. The field operator $\hat{\psi}_g(z)$ annihilates a ground state atom at position $z$. The second term in Eq.~(\ref{eq:1}) describes the dispersive atom-light coupling after the adiabatic elimination of the excited state. The  cavity Hamiltonian is given by $\hat{H}_{\mathrm{cav}}/\hbar=-\Delta_{c}\hat{a}^{\dagger}\hat{a}+\eta_{p}\hat{a}^{\dagger}+\eta_{p}^{*}\hat{a}$. 

The eigenstates of $\hat H_0$ are the so-called Wannier-Stark (WS) states. In the tight-binding limit, the wave function for a WS state centered at lattice site $n$ takes the form of $\phi_{n}(z)=\sum_{m} \mathcal{J}_{m-n}\left(2 J_{0}/Mga_{l}\right) w\left(z-m a_{l}\right)$~\cite{gluck2002wannier,dpt3}, which is a superposition of localized ground-band Wannier functions $w(z)$ [See Fig.~\ref{fig:setup}(b)]. Here $\mathcal{J}_n$ is the Bessel function of the first kind, $J_0/\hbar$ is the nearest-neighbour tunneling rate, and $a_l=\lambda_l/2$ is  the lattice spacing. 
The eigenenergy of $|\phi_n\rangle$ is $n\hbar\omega_B$, where $\omega_B=Mg a_l / \hbar$ is the Bloch frequency and  $T_B=2\pi/\omega_B$ the corresponding Bloch period.
We expand the field operator in the WS basis, $\hat{\psi}_g(z)=\sum_n \hat{c}_n \phi_n(z)$, where the operator $\hat{c}_n$ annihilates an atom in  the WS state $\phi_n$. In this basis, Eq.~\eqref{eq:1} can be rewritten as
\begin{equation}
  \hat{H}=\hat{H}_{\mathrm{cav}}+\frac{\hbar\mathcal{G}_0^{2}}{\Delta_{0}}\hat{a}^{\dagger}\hat{a}\hat{N}_{\mathrm{eff}}+\hbar\omega_{B}\sum_{n}n\hat{c}_{n}^{\dagger}\hat{c}_{n},
\label{eq:2}
\end{equation}
where
\begin{equation}
   \hat{N}_{\mathrm{eff}}=\sum_{m,n}J_{m,n}\hat{c}_{m}^{\dagger}\hat{c}_{n}.
   \label{eq:3} 
\end{equation}
Here, $J_{m,n}=\int dz \phi_m(z) \phi_n(z) \sin^2(k_c z)$ describes the overlap between the WS states $\phi_m$, $\phi_n$ weighted by the cavity field mode function. $\hat{N}_{\mathrm{eff}}$ can be understood as the effective number of atoms coupled to the cavity, which are responsible for generating  a frequency shift $\mathcal{G}^2_0 N_{\mathrm{eff}}/\Delta_0$ on the cavity resonance, where $N_{\mathrm{eff}}=\langle \hat{N}_{\mathrm{eff}}\rangle$.
This dispersive term allows us to either perform non-destructive probing  or   many-body control of the atomic motion, depending on  the operating  parameter regime. 

%In the case of incommensurate lattice and cavity wavelengths ($2\lambda_l/\lambda_c$ is a non-integer) as we discussed in this paper, the off-diagonal terms of $J_{m,n}$ ($m\neq n$) becomes non-zero, which generates effective hopping between WS states.

Assuming the cavity field adiabatically follows the atomic motion, which is valid since the cavity field dynamics ($\Delta_c\sim$ MHz) is much faster than the time evolution of the atomic ﬁeld  ($\omega_B\sim$ kHz), one can replace the cavity field operator by  $\hat{a}\approx\eta_p/(\Delta_c-\mathcal{G}^2_0 \hat{N}_{\mathrm{eff}}/\Delta_0)$. This leads to the following effective atom-only Hamiltonian~\cite{Note1},
\begin{equation}
    \hat{H}_{\mathrm{eff}}/\hbar=\omega_{B}\sum_{n}n\hat{c}_{n}^{\dagger}\hat{c}_{n}+\hat{V}_{\mathrm{cav}}(\hat{N}_{\mathrm{eff}}),
    \label{eq:heff}
\end{equation}
where $\hat{V}_\mathrm{cav}(\hat{N}_{\mathrm{eff}})=-(VN/\beta)/(1+\beta \hat{N}_{\mathrm{eff}}/N)$ is the cavity-induced potential depending on the atomic motion. Here, $V=\mathcal{G}^2_0|\eta_p|^2/(\Delta_c^2\Delta_0)$ is the maximum AC Stark shift on the atoms introduced by the bare cavity mode, $\beta=-N\mathcal{G}_0^2/(\Delta_0\Delta_c)$ is the ratio between the maximum cavity shift and the bare cavity detuning.
We assume $\beta>0$ ($\Delta_0$ and $\Delta_c$ have opposite signs) to avoid hitting a cavity resonance.

\section{Single-particle dynamics} First we consider the simplest case where the cavity is used as a probe and does not affect the single-particle dynamics set by $\hat{H}_0$, valid in the regimes $V \ll \omega_B$.
We consider the case where atoms are initially loaded in an almost localized WS state in a deep lattice at sites $n$ minimally coupled to the cavity ($k_c n a_l/\pi=r$ with $n,r\in\mathbb{Z}$). Then we suddenly quench the lattice depth to a shallow depth, and the atoms start hopping to the nearest-neighbour sites [see Fig.~\ref{fig:setup}(a)]. Since the  initially localized state corresponds to a  superposition of WS states of  the shallow lattice [see Fig.~\ref{fig:setup}(b)], after the quench, each WS state acquires a phase that evolves at a rate set by $\omega_B$. The interference of different WS states induces  tunneling away from the initially populated site, resulting in coherent breathing behavior at the BO frequency $\omega_B$.

%We  assume an initial state corresponding to  a localized Wannier Orbital in a  lattice with depth $V_0$ at site $n=0$ (not couple to the cavity). Then we quench  the system to a shallow lattice. As shown in \ref{fig:setup}(a). On the basis of WS the initial state evolves as $\left|\psi\right\rangle_{t}=\sum_n C_n \exp(-in\omega_Bt)$,
%  which generates tunneling away the initially populated site at a rate set by the Bloch oscillation frequency $\omega_B$.
To probe the BO, we use the fact that atoms at different sites coupled differently to the cavity. Therefore  tunneling out and back into  the initial site  leads to a periodic oscillation in  $N_{\mathrm{eff}}(t)$ at frequency $\omega_B$ as shown in Fig.~\ref{fig:setup}(d), which can be measured by tracking the cavity frequency shift $\mathcal{G}^2_0 N_{\mathrm{eff}}(t)/ \Delta_0$. 
Note that a technique to initially prepare atoms at lattice sites with low initial coupling to the cavity mode has been demonstrated in~\cite{wu2021site}.
Instead of an initially localized state, we can also use amplitude modulation of the lattice depth~\cite{bo9} to prepare a superposition of WS states. In this case a  similar  behavior can be observed as detailed in~\cite{Note1}.

For the numerical simulations throughout this letter,  we  consider the case of  $^{87}\mathrm{Rb}$ atoms with cavity wavelength $\lambda_c=780$ nm and lattice wavelength $\lambda_l=532$ nm.
However, the  discussion can be easily adapted to other type of atoms discussed in \cite{Note1}.

%%%%%%%%%%%%%%%%%%%%%%%%%%%%%%%%%%%%%%%%%%%%%%%%%%%%%%%%%%%%%%%
\begin{figure}[t]
	\centering
	\includegraphics[width=1\columnwidth]{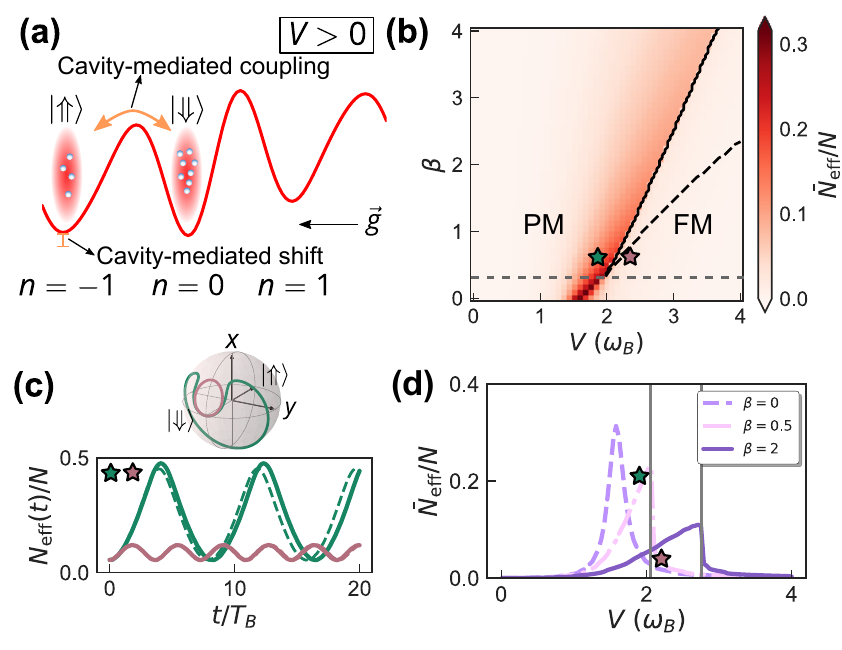}
	\caption{\label{fig:dpt} Dynamical Phase Transition (DPT) in the deep lattice regime ($V_0=20E_R$). 
	(a) For the case of $V>0$ we define an effective spin-1/2 degrees of freedom: $\left|\Uparrow\right\rangle$ ($\left|\phi_{-1}\right\rangle$) and $\left|\Downarrow\right\rangle$ ($\left|\phi_{0}\right\rangle$). The cavity-mediated interactions generate energy shifts to balance the potential energy of these two sites (red curve), as well as dynamical couplings between them (orange arrow).  
	(b) Phase diagram of the DPT determined by the long-time average $\bar{N}_{\mathrm{eff}}/N$. The phase boundary separating the paramagnetic (PM) and ferromagnetic (FM) phase is predicted by the full model (solid line) and LMG model (black dashed line). The smooth crossover regime is below the gray dashed line.
	(c) Mean-field dynamics with $V=1.9\omega_B,\beta=0.5$ (green) and $V=2.2\omega_B,\beta=0.5$ (red). The upper panel shows the mean-field trajectories on the Bloch sphere, and the lower panel displays the normalized signal $N_{\mathrm{eff}}(t)/N$. The solid (dashed) line show predictions of the full (LMG) model respectively.  
    (d) Horizontal cut of the phase diagram in (b) for $\beta=0$ (solid), $\beta=0.5$ (dashed), $\beta=2$ (dot dashed).
    }
\end{figure}
%%%%%%%%%%%%%%%%%%%%%%%%%%%%%%%%%%%%%%%%%%%%%%%%%%%%%%%%%%%%%%%

\section{Deep lattice regime} The interplay between single-particle atomic motion and cavity-mediated interactions occurs if $V\sim\omega_B$.
Here we focus on the deep lattice regime ($V_0=20E_R$) where WS states are almost localized at individual lattice sites. If atoms are  prepared  at site $n=0$, and $V>0$, the differential cavity induced  AC Stark shift  
(first order in $\beta$ in the limit $\beta\ll 1$) between the $n=0$ and $n=-1$ sites $\propto V(J_{0,0}-J_{-1,-1})$ can compensate for their energy difference $ \hbar \omega_B$ as shown in Fig.~\ref{fig:dpt}(a), restoring  tunnelling  between these two sites.
Since the atomic motion is restricted to take place between these two states, we map them to an effective spin $1/2$ degree of freedom: $\hat{c}_{-1}$ as $\hat{c}_{\Uparrow}$, $\hat{c}_{0}$ as $\hat{c}_{\Downarrow}$ as well as the spin operators $\hat{S}_z=(\hat{c}_{\Uparrow}^{\dagger}\hat{c}_{\Uparrow} - \hat{c}_{\Downarrow}^{\dagger}\hat{c}_{\Downarrow})/2$, $\hat{S}_x=(\hat{c}_{\Uparrow}^{\dagger}\hat{c}_{\Downarrow} + \hat{c}_{\Downarrow}^{\dagger}\hat{c}_{\Uparrow})/2$.
% $\hat c_{-1}^\dagger\left|0\right\rangle$ as $\left|\Uparrow\right\rangle$, and $\hat c_{0}^\dagger\left|0\right\rangle$ as $\left|\Downarrow\right\rangle$.
%To study the interplay between cavity mediated interactions and atomic motion we prepare an initial state $\left|\phi_0\right\rangle$ and let it evolve keeping the drive on. 
%In the deep lattice regime   ($V_0=20E_R$) while the bare  WS state is almost  localized  at site $n=0$, the cavity mediated interactions can generate  strong coupling between $\left|\phi_0\right\rangle$ and $\left|\phi_{-1}\right\rangle$ or $\left|\phi_0\right\rangle$ and $\left|\phi_{1}\right\rangle$ depending on the sign of $V$. 
%More explicitly, in this deep lattice limit where the dynamics is restricted to take place between atoms in the $\left|\phi_0\right\rangle$ and $\left|\phi_{-1}\right\rangle$ states, then population in other WS states can be neglected and this two states can be mapped to an effective spin $1/2$ degrees of freedom: $\left|\phi_{-1}\right\rangle$ as $\left|\Uparrow\right\rangle$ and $\left|\phi_{0}\right\rangle$ as $\left|\Downarrow\right\rangle$ as shown in Fig. \ref{fig:dpt}(a). 
% We can then define the collective spin operators $\hat{S}_{x,y,z}=\sum_{m,n}\hat{c}_m^{\dagger} \sigma^{m,n}_{x,y,z} \hat{c}_n/2$, where $\sigma^{m,n}_{x,y,z}$ are the matrix elements of the Pauli operators and $m,n\in \{\Uparrow ,\Downarrow\}$. 
Thus we have $\hat{N}_{\mathrm{eff}}=2(\Delta_{-1}\hat{S}_z+\Omega_{-1} \hat{S}_x)+N\bar{\omega}$, where $\Delta_{-1}=(J_{-1,-1}-J_{0,0})/2$, $\Omega_{-1}=J_{-1,0}$, and $\bar{\omega}=(J_{-1,-1}+J_{0,0})/2$.

In the limit of $\beta \ll 1$, one can expand $\hat{H}_{\mathrm{eff}}$ [Eq.~(\ref{eq:heff})] in a power series of $\beta$, and keep only the leading order terms.
The Hamiltonian simplifies to,
\begin{equation}
    \hat{H}_{\mathrm{eff}}/\hbar \approx -\omega_B\hat{S}_z + V \hat{N}_{\mathrm{eff}} - \frac{V\beta}{N}\hat{N}_{\mathrm{eff}}^2.
    \label{eq:5}
\end{equation}
% \begin{equation}
%     \begin{aligned}
%     \hat{H}_{\mathrm{eff}}/\hbar&\approx -\omega_B\hat{S}_z + \frac{(1-\beta\bar{\omega})V}{(1+\beta \bar{\omega})^3} (2\Delta_{-1}\hat{S}_z+2\Omega_{-1} \hat{S}_x)\\
%     &\frac{(-2\beta+\beta^2\bar{\omega})V}{N(1+\beta \bar{\omega})^4}(2\Delta_{-1}\hat{S}_z+2\Omega_{-1} \hat{S}_x)^2.
%     \end{aligned}
%     \label{eq:5}
% \end{equation}
%\begin{equation}
%\hat{H}=-\omega_B \hat{S}_z + 2 V_{\mathrm{cav}}(\beta,N_{\mathrm{eff}}) (\frac{\Delta_{-1}}{2} \hat{S}_z + \Omega_{-1} \hat{S}_x),  \label{eq:5}
%\end{equation}
%here $N_{\mathrm{eff}}=\Delta_{-1}  \left\langle \hat{S}_z \right\rangle + 2  \Omega_{-1} \left\langle \hat{S}_x \right\rangle + N \bar{\omega}$ and $\bar{\omega}=(J_{-1,-1}+J_{0,0})/2$. We define $J_{n,n}-J_{0,0}=\Delta_n$, $J_{n,n+1}=\Omega_n$ to simplify the expression.
This approximated Hamiltonian [Eq.~(\ref{eq:5})] is equivalent to the LMG model~\cite{dpt1,ma2011quantum,pezze2018quantum,li2022improving,Note1,dpt2,dpt3,dpt4},  $H_\mathrm{LMG}=\chi \hat{\tilde{S}}_z^2  + \tilde {\Omega} \hat{\tilde S}_x - \tilde{\delta} \hat{\tilde S}_z$, by a rotation along the $y$-axis of the Bloch sphere, $\hat{\tilde S}_\alpha=\hat R^\dagger \hat{S}_\alpha R$, where $\hat R=\exp(i\theta \hat S_{y})$, and $\text{\ensuremath{\tan\theta}}=\Delta_{-1}/\Omega_{-1}$~\cite{Note1}, which enables fast entanglement state generation under particular choice of $\chi,
\tilde{\Omega},\tilde{\delta}$~\cite{ma2011quantum,pezze2018quantum,li2022improving}.
%where $\hat{\tilde S}_\alpha=\hat R^\dagger \hat{\tilde S}_\alpha R$ with $R$ the rotation operator 
%along y-axis of the Bloch sphere ($\hat R=\exp(i\theta \hat S_{y})$ and $\text{\ensuremath{\tan\theta}}=\frac{\Delta_{-1}}{2\Omega_{-1}}$, See SM). 
The LMG model features a DPT from a dynamical ferromagnetic (FM) to a dynamical paramagnetic phase (PM), signaled by a sharp change in the behavior of the long-time average of the excitation fraction~\cite{dpt1,dpt2}. 
In our model [Eq.~(\ref{eq:heff})], the long-time average of the signal $\overline{N}_{\mathrm{eff}}/N=\lim_{T\rightarrow\infty}\int_0^T dt N_\mathrm{eff}(t)/(TN)$ plays the role of the dynamical order parameter. 
We also show that the DPT exists in our model [Eq.~(\ref{eq:heff})] even beyond the $\beta \ll 1$ limit as we discuss below.

To find the DPT, we solve the mean-field equations of motion for $s_{x,y,z}=2\langle \hat{S}_{x,y,z}\rangle/N$.
%as $\dot{\mathbf{s}}=\mathbf{b}(\mathbf{s})\times\mathbf{s}$,  and $\mathbf{b}(\mathbf{s})=\{\tilde\Omega,0,-\tilde \delta+2\chi \mathbf{s}_z\} $ (** check sign).
Such non-linear dynamics can be further reduced to $(\dot{N}_\mathrm{eff}/N)^2+f(N_\mathrm{eff}/N)=0$ with $f(J_{0,0})=0$, and we can associate the DPT with an abrupt change in the number of real roots of the effective potential $f(N_\mathrm{eff}/N)$~\cite{Note1}. 
This leads to the distinct dynamical behaviors of $N_{\mathrm{eff}}/N$ tuned by varying $V$ and $\beta$ as shown in Fig.~\ref{fig:dpt}(b,c,d). 
When the dynamics are dominated by interaction effects, the system is in the FM phase where the Bloch vector features small oscillations around the south pole, also shown as small amplitude oscillations in $N_{\mathrm{eff}}(t)/N$.
This phase is separated by a DPT to a PM phase where the Bloch vector exhibits large excursions around the Bloch sphere, also shown as large amplitude oscillations in  $N_{\mathrm{eff}}(t)/N$. 
For $\beta<0.32$~\cite{Note1}, the DPT transforms into a smooth crossover and the dynamics becomes dominated by single-particle tunneling processes. 
The dynamical phase boundary is plotted in Fig.~\ref{fig:dpt}(b) with the full model (solid line) and the LMG model (dashed line). The LMG model is unable to capture the phase boundary beyond the $\beta\ll 1$ limit.
%In Fig. \ref{fig:dpt}(d) cuts of the phase diagram at  fixed $\beta$ are shown where  $V$ is scan between $0$ and $4\omega_B$. They include  $\beta=0$, $0.5$, and $2$ curves. %The simulations are done for the 31 WS states, whereas the dynamics only happens between two WS states as expected.

%%%%%%%%%%%%%%%%%%%%%%%%%%%%%%%%%%%%%%%%%%%%%%%%%%%%%%%%%%%%%%%
\begin{figure}[t]
	\centering
	\includegraphics[width=1\columnwidth]{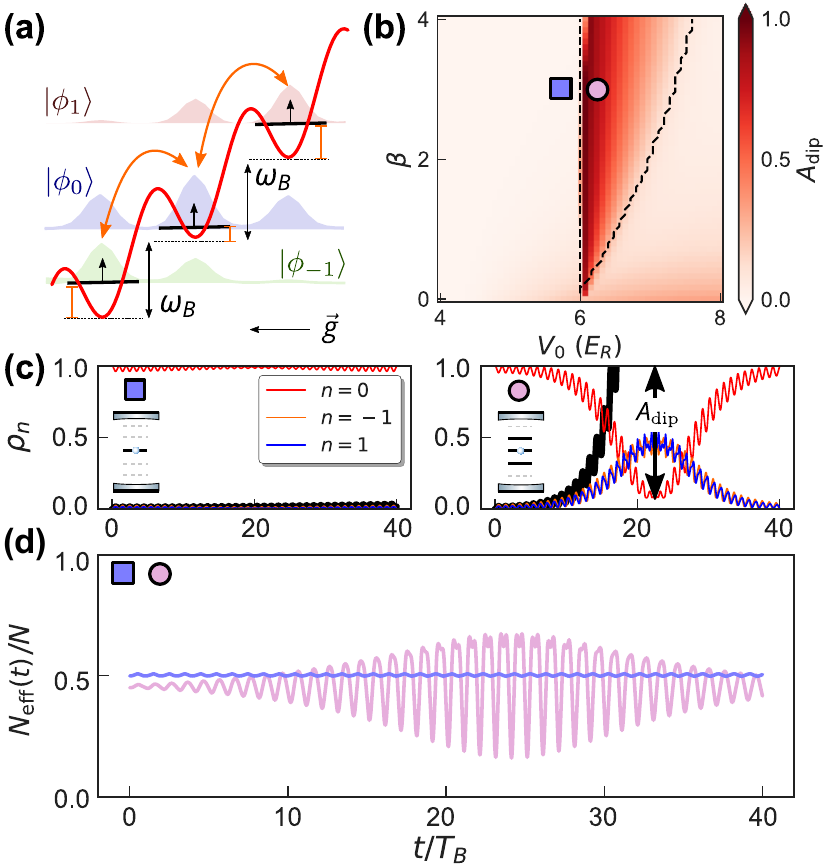}
	\caption{\label{fig:pair} Cavity-mediated amplification of Bloch oscillations in the shallow lattice regime. (a) The red lines show  the gravity plus optical lattice potential. Around $V_0 \approx 6E_R$, WS states can extend to the nearest-neighbour lattice sites. The orange vertical lines represent the cavity-induced onsite shift of the energy levels  and the orange arrows illustrate the cavity-mediated tunneling process shown in Eq.~\eqref{eq:6}. 
	(b) Transition between amplification regime and normal regime indicated by $A_{\mathrm{dip}}=1-\mathrm{min}\{\rho_0\}$. $V$ is fixed to be $2\omega_B$. The black dashed line shows the predicted boundary from UPA. 
	(c) Mean-field dynamics of $\rho_n$ with initial state $\left|\phi_0\right\rangle$ and $V=2\omega_B,\beta=3$. Nearly no dynamics happen in the left panel (purple square, $V_0=5.8E_R$) while large population transfer to $\left|\phi_1\right\rangle$ and $\left|\phi_{-1}\right\rangle$ (pink circle, $V_0=6.2E_R$) is observed  in the right panel. (d) Mean-field simulations for the normalized signal $N_{\mathrm{eff}}(t)/N$ for the same  parameters described in (c). The purple line stays almost constant while the pink line signals the cavity enhancement of the BO. 
	%The inset shows the Fourier transformation for the obtained $\bar{N}_{\mathrm{eff}}(t)/N$ (for the pink line), and the result shows the signal has the dominant component with the Bloch frequency $\omega_B$.
	}
\end{figure}
%%%%%%%%%%%%%%%%%%%%%%%%%%%%%%%%%%%%%%%%%%%%%%%%%%%%%%%%%%%%%%%

\section{Shallow lattice regime}
In a shallow lattice, the WS states extend over a few adjacent lattice sites. In this case, one can obtain a  significant suppression of differential AC Stark shifts generated by the cavity  by operating near the so-called magic lattice depth ($V_0=6E_R$ for the Rb parameters we use)~\cite{dpt3}, where $J_{n,n}$ is nearly a constant and the energy difference between nearest-neighbour WS states is roughly $\hbar\omega_B$ [see Fig.~\ref{fig:pair}(a)]. Thus the dynamics features BO even in the presence of strong cavity-mediated interactions. 
In fact, after preparing the atoms in the  WS state $\left|\phi_0\right\rangle$ and thus in an  eigenstate of the single particle Hamiltonian, one can observe the generation and amplification of BO due to cavity-mediated interactions in a window around the magic depth as shown in Fig.~\ref{fig:pair}(b,c,d). 
Since the  short-time dynamics occurs mainly between the WS states centered at $n=0,\pm1$, we can concentrate only  on these states  and  simplify the dynamics via the  undepleted pump approximation (UPA):  To the leading order, one can replace the operators for the initially occupied states as c-numbers, $\hat{c}_0, \hat{c}_0^{\dag}\sim \sqrt{N}$, and keep the operators for unoccupied states ($\hat{c}_{\pm 1}, \hat{c}_{\pm 1}^{\dag}$) to the second order while absorbing  the linear term generated by single-particle tunneling via a displacement of a coherent state, $\hat{c}_{\pm 1}=\alpha_{\pm 1}+\hat{c}'_{\pm 1}$.
In this way, $\hat{H}_{\mathrm{eff}}$ [Eq.~(\ref{eq:heff})] simplifies into a quadratic form~\cite{Note1},
\begin{equation}
\begin{aligned}
  \hat{H}_{\mathrm{eff}}/\hbar &\approx  \omega_B (\hat{c}'^\dagger_{1} \hat{c}'_1  - \hat{c}'^\dagger_{-1} \hat{c}'_{-1}) + V_1 \Delta (\hat{c}'^\dagger_{1} \hat{c}'_1 + \hat{c}'^\dagger_{-1} \hat{c}'_{-1}) \\
  & + V_2 N \Omega^2 (\hat{c}'^\dagger_{1} + \hat{c}'_1 - \hat{c}'^\dagger_{-1} - \hat{c}'_{-1})^2,
\end{aligned}
    \label{eq:6}
\end{equation}
with the expansion coefficient $V_1$, $V_2$ in~\cite{Note1} and $\Delta=\Delta_1\approx\Delta_{-1}$, $\Omega=\Omega_1\approx -\Omega_{-1}$.

We analyze the exact dynamics of Eq.~(\ref{eq:6}) via the  Bogoliubov-de Gennes method, in which the Heisenberg equation of motion for operators $\hat{C}=(\hat{c}'_1,\hat{c}'_{-1},\hat{c}_1^{\prime\dag},\hat{c}_{-1}^{\prime\dag})^{T}$ takes the form $i\partial_t \hat{C}=\mathcal{H}_{\mathrm{BdG}}\hat{C}$.
The matrix $\mathcal{H}_{\mathrm{BdG}}$ can have either real or complex eigenvalues, which leads to distinct dynamical behaviors as shown in Fig.~\ref{fig:pair}(c).  When  all the eigenvalues are real (normal regime), the populations $\rho_0$ and $\rho_{\pm1}$,  with  $\rho_{n}=\langle \hat{c}_n^{\dag}\hat{c}_n\rangle$,
feature stable small amplitude oscillations; on the other hand when  all the eigenvalues are complex, then $\rho_{\pm1}$ feature  an exponential growth associated with the
correlated pair production of atoms at WS centered at $n=\pm 1$,  which leads to the amplification of the BO signal until UPA breaks down.  The transition between the real and complex eigenvalues of $\mathcal{H}_{\mathrm{BdG}}$ is marked by dashed lines in Fig.~\ref{fig:pair}(b).

To quantify the population transfer, we define $A_{\mathrm{dip}}=1-\mathrm{min}\{\rho_0\}$ with $\mathrm{min}\{\rho_0\}$ as the minimum of $\rho_0$ during $t\in[0,40T_B]$. A large $A_{\mathrm{dip}}$ signals efficient population transfer. In Fig.~\ref{fig:pair}(b), we show $A_{\mathrm{dip}}$ as a function of the lattice depth $V_0$ and the cavity parameter $\beta$. The region of amplified BO lies  within the  two dashed boundaries.  The  left boundary  is fixed at the magic lattice depth ($V_0=6E_R$)  and the right boundary  pushes to larger $\beta$ as $V_0$ increases. Inside the  amplification region   $A_{\mathrm{dip}}\neq 0 $, while outside  $A_{\mathrm{dip}}\approx 0$.
The evolution of $N_\mathrm{eff}(t)/N$ is shown in Fig. \ref{fig:pair}(d), where the enhanced population transfer induced by the cavity-mediated interactions lead to the growth of the BO amplitude in the amplification regime.

\section{Experimental consideration} The predicted behavior should be achievable in state-of-the-art cavity QED systems with $N\sim 10^4$ $^{87}\mathrm{Rb}$ atoms. We focus on the unitary dynamics in this letter while the main decoherence sources come from cavity loss and spontaneous emission from the excited states.
The cavity loss generates collective dephasing processes at a rate $V\beta \kappa/\Delta_c$ and spontaneous emission generates off-resonant photon scattering processes at a rate $V\gamma / \Delta_0$, where $\gamma$ is the spontaneous emission rate. 
For an optical cavity with cooperativity $C=4\mathcal{G}_0^2/\gamma \kappa\sim 0.5$,  $\kappa/\Delta_c\sim 0.05$, $\gamma/\Delta_0\sim 0.01$, one obtains  negligible dissipation within experimentally relevant  time scales ($\sim50$ BO periods). 
Our scheme does not require BEC while utilizes site-selection to prepare the initial state, which is robust to the radial thermal noise up to $T\sim1\mu$K~\cite{Note1}.
% The ultracold temperature requirement for initial states is significantly suppressed in our protocol: as shown in~\cite{Note1}, the many-body phenomena remains unaffected with radial temperature $T\sim1\mu$K.
Contact interactions between atoms can also be ignored for the dilute quantum gas used here ($\sim 50$ atoms per site).
%for $\gamma\ll\Delta_0$.  The ratio $\beta$ can be repressed as $- N C/4\delta \times \gamma/\Delta_0$ and tuned to be $O(1)$, where $C=4\mathcal{G}_0^2/\gamma \kappa$ for cooperativity and $\delta=\Delta_c/\kappa$ for the ratios between cavity detuning and linewidth. Thus large particle number $N$ and cooperativity $C$ are favored for $\gamma\ll\Delta_0$, $\Delta_c\ll\kappa$ while maintain $\beta$. As a example, with $N\sim2\times10^4$, $C\sim0.5$ and $\delta\sim20$, we can obtain $\gamma/\Delta_0=0.01$ and $\beta=1.25$.
%For the particular case of $^{87}\mathrm{Rb}$ atoms with a lattice wavelength ($D_2$ transtion) $\lambda_l=532$ nm and a cavity wavelength $\lambda_c=780$ nm will give the magic lattice depth $V_0=6E_R$.
Moreover, our model can be realized with other species of alkali atoms ($D_2$ transition) and alkaline earth atoms (${}^1S_0\rightarrow {}^3P_1$ transition) with appropriate choices of lattice wavelength~\cite{Note1}.
In particular, contact interactions can be further suppressed using $^{88}\mathrm{Sr}$ atoms featuring negligible scattering lengths or any type of fermionic atoms interacting only via the $p$-wave channel.  
%, such as boson: $^{87}\mathrm{Sr}$ and fermion: $^{88}\mathrm{Sr}$ and $^{171}\mathrm{Yb}$. For $^{171}\mathrm{Yb}$ atoms with $\lambda_c=556$nm ($^1S_0\rightarrow ^3P_1$ transition) and $\lambda_l=413$nm, the magic wavelength is around $3.2E_R$. $^{171}\mathrm{Yb}$ atoms also features negligible scattering length thus the contact interaction can be ignored. 
%We only focus on the unitary dynamics with the current experiment parameters and assume dissipation don't play the significant roles during the dynamics.  

\section{Conclusion and outlook} In summary, we proposed a scheme to perform many-body control of atomic BO in an optical cavity.
%and discuss the rich many-body dynamics introduced by the cavity-mediated long-range interactions, such as dynamical phase transition as well as amplification of BO in different parameter regimes. 
Our work opens new possibilities for  Hamiltonian engineering in many-body systems by taking advantage of  the interplay between atomic motion, gravity and cavity-mediated interactions. For example, although so far we only focused on a single internal level, by  including more levels  and  more cavity modes, it should be possible  to engineer dynamical self-generated couplings between WS states via cavity-mediated interactions, which could be used  to study  dynamical gauge field~\cite{Helmut2,Helmut3,rosa2022} in a synthetic ladder without the overhead of Raman beams.  Furthermore, although most of the calculations so far have been limited to regimes where the mean-field dynamics are a good description of the system,  by loading the atoms in 2D or 3D lattice, one should be able to increase the role of beyond mean-field effects and enter the regimes where quantum correlations dominate the dynamics. 

\begin{acknowledgments}
We thank Tianrui Xu and Tobias Bothwell for critical reading of the manuscript, and we thank Helmut Ritsch for useful discussions. 
This work is supported by the AFOSR Grant No. FA9550-18-1-0319, by the DARPA (funded via ARO) Grant No. W911NF-16-1-0576, the ARO single investigator Grant No. W911NF-19-1-0210, the NSF PHY1820885, NSF JILA-PFC PHY-1734006 and NSF QLCI-2016244 grants, by the DOE Quantum Systems Accelerator (QSA) grant and by NIST.
\end{acknowledgments}

\bibliography{apssamp}% Produces the bibliography via BibTeX.

\providecommand{\noopsort}[1]{}\providecommand{\singleletter}[1]{#1}%
\begin{thebibliography}{38}%
\makeatletter
\providecommand \@ifxundefined [1]{%
 \@ifx{#1\undefined}
}%
\providecommand \@ifnum [1]{%
 \ifnum #1\expandafter \@firstoftwo
 \else \expandafter \@secondoftwo
 \fi
}%
\providecommand \@ifx [1]{%
 \ifx #1\expandafter \@firstoftwo
 \else \expandafter \@secondoftwo
 \fi
}%
\providecommand \natexlab [1]{#1}%
\providecommand \enquote  [1]{``#1''}%
\providecommand \bibnamefont  [1]{#1}%
\providecommand \bibfnamefont [1]{#1}%
\providecommand \citenamefont [1]{#1}%
\providecommand \href@noop [0]{\@secondoftwo}%
\providecommand \href [0]{\begingroup \@sanitize@url \@href}%
\providecommand \@href[1]{\@@startlink{#1}\@@href}%
\providecommand \@@href[1]{\endgroup#1\@@endlink}%
\providecommand \@sanitize@url [0]{\catcode `\\12\catcode `\$12\catcode
  `\&12\catcode `\#12\catcode `\^12\catcode `\_12\catcode `\%12\relax}%
\providecommand \@@startlink[1]{}%
\providecommand \@@endlink[0]{}%
\providecommand \url  [0]{\begingroup\@sanitize@url \@url }%
\providecommand \@url [1]{\endgroup\@href {#1}{\urlprefix }}%
\providecommand \urlprefix  [0]{URL }%
\providecommand \Eprint [0]{\href }%
\providecommand \doibase [0]{http://dx.doi.org/}%
\providecommand \selectlanguage [0]{\@gobble}%
\providecommand \bibinfo  [0]{\@secondoftwo}%
\providecommand \bibfield  [0]{\@secondoftwo}%
\providecommand \translation [1]{[#1]}%
\providecommand \BibitemOpen [0]{}%
\providecommand \bibitemStop [0]{}%
\providecommand \bibitemNoStop [0]{.\EOS\space}%
\providecommand \EOS [0]{\spacefactor3000\relax}%
\providecommand \BibitemShut  [1]{\csname bibitem#1\endcsname}%
\let\auto@bib@innerbib\@empty
%</preamble>
\bibitem [{\citenamefont {Bloch}(1929)}]{bloch1929}%
  \BibitemOpen
  \bibfield  {author} {\bibinfo {author} {\bibfnamefont {F.}~\bibnamefont
  {Bloch}},\ }\href {https://link.springer.com/article/10.1007/BF01339455}
  {\bibfield  {journal} {\bibinfo  {journal} {Zeitschrift f{\"u}r physik}\
  }\textbf {\bibinfo {volume} {52}},\ \bibinfo {pages} {555} (\bibinfo {year}
  {1929})}\BibitemShut {NoStop}%
\bibitem [{\citenamefont {Waschke}\ \emph {et~al.}(1993)\citenamefont
  {Waschke}, \citenamefont {Roskos}, \citenamefont {Schwedler}, \citenamefont
  {Leo}, \citenamefont {Kurz},\ and\ \citenamefont {K{\"o}hler}}]{bo0}%
  \BibitemOpen
  \bibfield  {author} {\bibinfo {author} {\bibfnamefont {C.}~\bibnamefont
  {Waschke}}, \bibinfo {author} {\bibfnamefont {H.~G.}\ \bibnamefont {Roskos}},
  \bibinfo {author} {\bibfnamefont {R.}~\bibnamefont {Schwedler}}, \bibinfo
  {author} {\bibfnamefont {K.}~\bibnamefont {Leo}}, \bibinfo {author}
  {\bibfnamefont {H.}~\bibnamefont {Kurz}}, \ and\ \bibinfo {author}
  {\bibfnamefont {K.}~\bibnamefont {K{\"o}hler}},\ }\href
  {https://journals.aps.org/prl/abstract/10.1103/PhysRevLett.70.3319}
  {\bibfield  {journal} {\bibinfo  {journal} {Physical review letters}\
  }\textbf {\bibinfo {volume} {70}},\ \bibinfo {pages} {3319} (\bibinfo {year}
  {1993})}\BibitemShut {NoStop}%
\bibitem [{\citenamefont {Dahan}\ \emph {et~al.}(1996)\citenamefont {Dahan},
  \citenamefont {Peik}, \citenamefont {Reichel}, \citenamefont {Castin},\ and\
  \citenamefont {Salomon}}]{dahan1996bloch}%
  \BibitemOpen
  \bibfield  {author} {\bibinfo {author} {\bibfnamefont {M.~B.}\ \bibnamefont
  {Dahan}}, \bibinfo {author} {\bibfnamefont {E.}~\bibnamefont {Peik}},
  \bibinfo {author} {\bibfnamefont {J.}~\bibnamefont {Reichel}}, \bibinfo
  {author} {\bibfnamefont {Y.}~\bibnamefont {Castin}}, \ and\ \bibinfo {author}
  {\bibfnamefont {C.}~\bibnamefont {Salomon}},\ }\href
  {https://journals.aps.org/prl/abstract/10.1103/PhysRevLett.76.4508}
  {\bibfield  {journal} {\bibinfo  {journal} {Physical Review Letters}\
  }\textbf {\bibinfo {volume} {76}},\ \bibinfo {pages} {4508} (\bibinfo {year}
  {1996})}\BibitemShut {NoStop}%
\bibitem [{\citenamefont {Anderson}\ and\ \citenamefont
  {Kasevich}(1998)}]{anderson1998macroscopic}%
  \BibitemOpen
  \bibfield  {author} {\bibinfo {author} {\bibfnamefont {B.~P.}\ \bibnamefont
  {Anderson}}\ and\ \bibinfo {author} {\bibfnamefont {M.~A.}\ \bibnamefont
  {Kasevich}},\ }\href
  {https://www.science.org/doi/10.1126/science.282.5394.1686} {\bibfield
  {journal} {\bibinfo  {journal} {Science}\ }\textbf {\bibinfo {volume}
  {282}},\ \bibinfo {pages} {1686} (\bibinfo {year} {1998})}\BibitemShut
  {NoStop}%
\bibitem [{\citenamefont {Ke{\ss}ler}\ \emph {et~al.}(2016)\citenamefont
  {Ke{\ss}ler}, \citenamefont {Klinder}, \citenamefont {Venkatesh},
  \citenamefont {Georges},\ and\ \citenamefont {Hemmerich}}]{Hemmerich1}%
  \BibitemOpen
  \bibfield  {author} {\bibinfo {author} {\bibfnamefont {H.}~\bibnamefont
  {Ke{\ss}ler}}, \bibinfo {author} {\bibfnamefont {J.}~\bibnamefont {Klinder}},
  \bibinfo {author} {\bibfnamefont {B.~P.}\ \bibnamefont {Venkatesh}}, \bibinfo
  {author} {\bibfnamefont {C.}~\bibnamefont {Georges}}, \ and\ \bibinfo
  {author} {\bibfnamefont {A.}~\bibnamefont {Hemmerich}},\ }\href
  {https://iopscience.iop.org/article/10.1088/1367-2630/18/10/102001}
  {\bibfield  {journal} {\bibinfo  {journal} {New Journal of Physics}\ }\textbf
  {\bibinfo {volume} {18}},\ \bibinfo {pages} {102001} (\bibinfo {year}
  {2016})}\BibitemShut {NoStop}%
\bibitem [{\citenamefont {Georges}\ \emph {et~al.}(2017)\citenamefont
  {Georges}, \citenamefont {Vargas}, \citenamefont {Ke{\ss}ler}, \citenamefont
  {Klinder},\ and\ \citenamefont {Hemmerich}}]{Hemmerich2}%
  \BibitemOpen
  \bibfield  {author} {\bibinfo {author} {\bibfnamefont {C.}~\bibnamefont
  {Georges}}, \bibinfo {author} {\bibfnamefont {J.}~\bibnamefont {Vargas}},
  \bibinfo {author} {\bibfnamefont {H.}~\bibnamefont {Ke{\ss}ler}}, \bibinfo
  {author} {\bibfnamefont {J.}~\bibnamefont {Klinder}}, \ and\ \bibinfo
  {author} {\bibfnamefont {A.}~\bibnamefont {Hemmerich}},\ }\href
  {https://journals.aps.org/pra/abstract/10.1103/PhysRevA.96.063615} {\bibfield
   {journal} {\bibinfo  {journal} {Physical Review A}\ }\textbf {\bibinfo
  {volume} {96}},\ \bibinfo {pages} {063615} (\bibinfo {year}
  {2017})}\BibitemShut {NoStop}%
\bibitem [{\citenamefont {Peden}\ \emph {et~al.}(2009)\citenamefont {Peden},
  \citenamefont {Meiser}, \citenamefont {Chiofalo},\ and\ \citenamefont
  {Holland}}]{peden2009}%
  \BibitemOpen
  \bibfield  {author} {\bibinfo {author} {\bibfnamefont {B.~M.}\ \bibnamefont
  {Peden}}, \bibinfo {author} {\bibfnamefont {D.}~\bibnamefont {Meiser}},
  \bibinfo {author} {\bibfnamefont {M.~L.}\ \bibnamefont {Chiofalo}}, \ and\
  \bibinfo {author} {\bibfnamefont {M.~J.}\ \bibnamefont {Holland}},\ }\href
  {\doibase 10.1103/PhysRevA.80.043803} {\bibfield  {journal} {\bibinfo
  {journal} {Phys. Rev. A}\ }\textbf {\bibinfo {volume} {80}},\ \bibinfo
  {pages} {043803} (\bibinfo {year} {2009})}\BibitemShut {NoStop}%
\bibitem [{\citenamefont {Venkatesh}\ \emph {et~al.}(2009)\citenamefont
  {Venkatesh}, \citenamefont {Trupke}, \citenamefont {Hinds},\ and\
  \citenamefont {O’Dell}}]{bo8}%
  \BibitemOpen
  \bibfield  {author} {\bibinfo {author} {\bibfnamefont {B.~P.}\ \bibnamefont
  {Venkatesh}}, \bibinfo {author} {\bibfnamefont {M.}~\bibnamefont {Trupke}},
  \bibinfo {author} {\bibfnamefont {E.~A.}\ \bibnamefont {Hinds}}, \ and\
  \bibinfo {author} {\bibfnamefont {D.~H.~J.}\ \bibnamefont {O’Dell}},\
  }\href {https://journals.aps.org/pra/abstract/10.1103/PhysRevA.80.063834}
  {\bibfield  {journal} {\bibinfo  {journal} {Physical Review A}\ }\textbf
  {\bibinfo {volume} {80}},\ \bibinfo {pages} {063834} (\bibinfo {year}
  {2009})}\BibitemShut {NoStop}%
\bibitem [{\citenamefont {Venkatesh}\ and\ \citenamefont
  {O'Dell}(2013)}]{venk2013}%
  \BibitemOpen
  \bibfield  {author} {\bibinfo {author} {\bibfnamefont {B.~P.}\ \bibnamefont
  {Venkatesh}}\ and\ \bibinfo {author} {\bibfnamefont {D.~H.~J.}\ \bibnamefont
  {O'Dell}},\ }\href {\doibase 10.1103/PhysRevA.88.013848} {\bibfield
  {journal} {\bibinfo  {journal} {Phys. Rev. A}\ }\textbf {\bibinfo {volume}
  {88}},\ \bibinfo {pages} {013848} (\bibinfo {year} {2013})}\BibitemShut
  {NoStop}%
\bibitem [{\citenamefont {Wu}\ \emph {et~al.}(2021)\citenamefont {Wu},
  \citenamefont {Greve}, \citenamefont {Luo},\ and\ \citenamefont
  {Thompson}}]{wu2021site}%
  \BibitemOpen
  \bibfield  {author} {\bibinfo {author} {\bibfnamefont {B.}~\bibnamefont
  {Wu}}, \bibinfo {author} {\bibfnamefont {G.~P.}\ \bibnamefont {Greve}},
  \bibinfo {author} {\bibfnamefont {C.}~\bibnamefont {Luo}}, \ and\ \bibinfo
  {author} {\bibfnamefont {J.~K.}\ \bibnamefont {Thompson}},\ }\href
  {https://doi.org/10.48550/arXiv.2104.01201} {\bibfield  {journal} {\bibinfo
  {journal} {arXiv:2104.01201}\ } (\bibinfo {year} {2021})}\BibitemShut
  {NoStop}%
\bibitem [{\citenamefont {Niederriter}\ \emph {et~al.}(2020)\citenamefont
  {Niederriter}, \citenamefont {Schlupf},\ and\ \citenamefont
  {Hamilton}}]{bo11}%
  \BibitemOpen
  \bibfield  {author} {\bibinfo {author} {\bibfnamefont {R.~D.}\ \bibnamefont
  {Niederriter}}, \bibinfo {author} {\bibfnamefont {C.}~\bibnamefont
  {Schlupf}}, \ and\ \bibinfo {author} {\bibfnamefont {P.}~\bibnamefont
  {Hamilton}},\ }\href
  {https://journals.aps.org/pra/abstract/10.1103/PhysRevA.102.051301}
  {\bibfield  {journal} {\bibinfo  {journal} {Physical Review A}\ }\textbf
  {\bibinfo {volume} {102}},\ \bibinfo {pages} {051301} (\bibinfo {year}
  {2020})}\BibitemShut {NoStop}%
\bibitem [{\citenamefont {Chu}\ \emph {et~al.}(2021)\citenamefont {Chu},
  \citenamefont {He}, \citenamefont {Thompson},\ and\ \citenamefont
  {Rey}}]{dpt3}%
  \BibitemOpen
  \bibfield  {author} {\bibinfo {author} {\bibfnamefont {A.}~\bibnamefont
  {Chu}}, \bibinfo {author} {\bibfnamefont {P.}~\bibnamefont {He}}, \bibinfo
  {author} {\bibfnamefont {J.~K.}\ \bibnamefont {Thompson}}, \ and\ \bibinfo
  {author} {\bibfnamefont {A.~M.}\ \bibnamefont {Rey}},\ }\href
  {https://journals.aps.org/prl/abstract/10.1103/PhysRevLett.127.210401}
  {\bibfield  {journal} {\bibinfo  {journal} {Physical Review Letters}\
  }\textbf {\bibinfo {volume} {127}},\ \bibinfo {pages} {210401} (\bibinfo
  {year} {2021})}\BibitemShut {NoStop}%
\bibitem [{\citenamefont {Buchleitner}\ and\ \citenamefont
  {Kolovsky}(2003)}]{bo1}%
  \BibitemOpen
  \bibfield  {author} {\bibinfo {author} {\bibfnamefont {A.}~\bibnamefont
  {Buchleitner}}\ and\ \bibinfo {author} {\bibfnamefont {A.~R.}\ \bibnamefont
  {Kolovsky}},\ }\href
  {https://journals.aps.org/prl/abstract/10.1103/PhysRevLett.91.253002}
  {\bibfield  {journal} {\bibinfo  {journal} {Physical review letters}\
  }\textbf {\bibinfo {volume} {91}},\ \bibinfo {pages} {253002} (\bibinfo
  {year} {2003})}\BibitemShut {NoStop}%
\bibitem [{\citenamefont {Kolovsky}(2003)}]{bo2}%
  \BibitemOpen
  \bibfield  {author} {\bibinfo {author} {\bibfnamefont {A.~R.}\ \bibnamefont
  {Kolovsky}},\ }\href
  {https://journals.aps.org/prl/abstract/10.1103/PhysRevLett.90.213002}
  {\bibfield  {journal} {\bibinfo  {journal} {Physical review letters}\
  }\textbf {\bibinfo {volume} {90}},\ \bibinfo {pages} {213002} (\bibinfo
  {year} {2003})}\BibitemShut {NoStop}%
\bibitem [{\citenamefont {Witthaut}\ \emph {et~al.}(2005)\citenamefont
  {Witthaut}, \citenamefont {Werder}, \citenamefont {Mossmann},\ and\
  \citenamefont {Korsch}}]{bo3}%
  \BibitemOpen
  \bibfield  {author} {\bibinfo {author} {\bibfnamefont {D.}~\bibnamefont
  {Witthaut}}, \bibinfo {author} {\bibfnamefont {M.}~\bibnamefont {Werder}},
  \bibinfo {author} {\bibfnamefont {S.}~\bibnamefont {Mossmann}}, \ and\
  \bibinfo {author} {\bibfnamefont {H.}~\bibnamefont {Korsch}},\ }\href
  {https://journals.aps.org/pre/abstract/10.1103/PhysRevE.71.036625} {\bibfield
   {journal} {\bibinfo  {journal} {Physical Review E}\ }\textbf {\bibinfo
  {volume} {71}},\ \bibinfo {pages} {036625} (\bibinfo {year}
  {2005})}\BibitemShut {NoStop}%
\bibitem [{\citenamefont {Schulte}\ \emph {et~al.}(2008)\citenamefont
  {Schulte}, \citenamefont {Drenkelforth}, \citenamefont {B{\"u}ning},
  \citenamefont {Ertmer}, \citenamefont {Arlt}, \citenamefont {Lewenstein},\
  and\ \citenamefont {Santos}}]{bo4}%
  \BibitemOpen
  \bibfield  {author} {\bibinfo {author} {\bibfnamefont {T.}~\bibnamefont
  {Schulte}}, \bibinfo {author} {\bibfnamefont {S.}~\bibnamefont
  {Drenkelforth}}, \bibinfo {author} {\bibfnamefont {G.~K.}\ \bibnamefont
  {B{\"u}ning}}, \bibinfo {author} {\bibfnamefont {W.}~\bibnamefont {Ertmer}},
  \bibinfo {author} {\bibfnamefont {J.}~\bibnamefont {Arlt}}, \bibinfo {author}
  {\bibfnamefont {M.}~\bibnamefont {Lewenstein}}, \ and\ \bibinfo {author}
  {\bibfnamefont {L.}~\bibnamefont {Santos}},\ }\href
  {https://journals.aps.org/pra/abstract/10.1103/PhysRevA.77.023610} {\bibfield
   {journal} {\bibinfo  {journal} {Physical Review A}\ }\textbf {\bibinfo
  {volume} {77}},\ \bibinfo {pages} {023610} (\bibinfo {year}
  {2008})}\BibitemShut {NoStop}%
\bibitem [{\citenamefont {Walter}\ \emph {et~al.}(2010)\citenamefont {Walter},
  \citenamefont {Schneble},\ and\ \citenamefont {Durst}}]{bo5}%
  \BibitemOpen
  \bibfield  {author} {\bibinfo {author} {\bibfnamefont {S.}~\bibnamefont
  {Walter}}, \bibinfo {author} {\bibfnamefont {D.}~\bibnamefont {Schneble}}, \
  and\ \bibinfo {author} {\bibfnamefont {A.~C.}\ \bibnamefont {Durst}},\ }\href
  {https://journals.aps.org/pra/abstract/10.1103/PhysRevA.81.033623} {\bibfield
   {journal} {\bibinfo  {journal} {Physical Review A}\ }\textbf {\bibinfo
  {volume} {81}},\ \bibinfo {pages} {033623} (\bibinfo {year}
  {2010})}\BibitemShut {NoStop}%
\bibitem [{\citenamefont {Meinert}\ \emph {et~al.}(2014)\citenamefont
  {Meinert}, \citenamefont {Mark}, \citenamefont {Kirilov}, \citenamefont
  {Lauber}, \citenamefont {Weinmann}, \citenamefont {Gr{\"o}bner},\ and\
  \citenamefont {N{\"a}gerl}}]{bo6}%
  \BibitemOpen
  \bibfield  {author} {\bibinfo {author} {\bibfnamefont {F.}~\bibnamefont
  {Meinert}}, \bibinfo {author} {\bibfnamefont {M.~J.}\ \bibnamefont {Mark}},
  \bibinfo {author} {\bibfnamefont {E.}~\bibnamefont {Kirilov}}, \bibinfo
  {author} {\bibfnamefont {K.}~\bibnamefont {Lauber}}, \bibinfo {author}
  {\bibfnamefont {P.}~\bibnamefont {Weinmann}}, \bibinfo {author}
  {\bibfnamefont {M.}~\bibnamefont {Gr{\"o}bner}}, \ and\ \bibinfo {author}
  {\bibfnamefont {H.-C.}\ \bibnamefont {N{\"a}gerl}},\ }\href
  {https://journals.aps.org/prl/abstract/10.1103/PhysRevLett.112.193003}
  {\bibfield  {journal} {\bibinfo  {journal} {Physical Review Letters}\
  }\textbf {\bibinfo {volume} {112}},\ \bibinfo {pages} {193003} (\bibinfo
  {year} {2014})}\BibitemShut {NoStop}%
\bibitem [{\citenamefont {Alberti}\ \emph {et~al.}(2010)\citenamefont
  {Alberti}, \citenamefont {Ferrari}, \citenamefont {Ivanov}, \citenamefont
  {Chiofalo},\ and\ \citenamefont {Tino}}]{bo9}%
  \BibitemOpen
  \bibfield  {author} {\bibinfo {author} {\bibfnamefont {A.}~\bibnamefont
  {Alberti}}, \bibinfo {author} {\bibfnamefont {G.}~\bibnamefont {Ferrari}},
  \bibinfo {author} {\bibfnamefont {V.~V.}\ \bibnamefont {Ivanov}}, \bibinfo
  {author} {\bibfnamefont {M.~L.}\ \bibnamefont {Chiofalo}}, \ and\ \bibinfo
  {author} {\bibfnamefont {G.~M.}\ \bibnamefont {Tino}},\ }\href
  {https://iopscience.iop.org/article/10.1088/1367-2630/12/6/065037} {\bibfield
   {journal} {\bibinfo  {journal} {New Journal of Physics}\ }\textbf {\bibinfo
  {volume} {12}},\ \bibinfo {pages} {065037} (\bibinfo {year}
  {2010})}\BibitemShut {NoStop}%
\bibitem [{\citenamefont {Masi}\ \emph {et~al.}(2021)\citenamefont {Masi},
  \citenamefont {Petrucciani}, \citenamefont {Ferioli}, \citenamefont
  {Semeghini}, \citenamefont {Modugno}, \citenamefont {Inguscio},\ and\
  \citenamefont {Fattori}}]{bo10}%
  \BibitemOpen
  \bibfield  {author} {\bibinfo {author} {\bibfnamefont {L.}~\bibnamefont
  {Masi}}, \bibinfo {author} {\bibfnamefont {T.}~\bibnamefont {Petrucciani}},
  \bibinfo {author} {\bibfnamefont {G.}~\bibnamefont {Ferioli}}, \bibinfo
  {author} {\bibfnamefont {G.}~\bibnamefont {Semeghini}}, \bibinfo {author}
  {\bibfnamefont {G.}~\bibnamefont {Modugno}}, \bibinfo {author} {\bibfnamefont
  {M.}~\bibnamefont {Inguscio}}, \ and\ \bibinfo {author} {\bibfnamefont
  {M.}~\bibnamefont {Fattori}},\ }\href
  {https://journals.aps.org/prl/abstract/10.1103/PhysRevLett.127.020601}
  {\bibfield  {journal} {\bibinfo  {journal} {Physical Review Letters}\
  }\textbf {\bibinfo {volume} {127}},\ \bibinfo {pages} {020601} (\bibinfo
  {year} {2021})}\BibitemShut {NoStop}%
\bibitem [{\citenamefont {Chu}\ \emph {et~al.}(2020)\citenamefont {Chu},
  \citenamefont {Will}, \citenamefont {Arlt}, \citenamefont {Klempt},\ and\
  \citenamefont {Rey}}]{dpt1}%
  \BibitemOpen
  \bibfield  {author} {\bibinfo {author} {\bibfnamefont {A.}~\bibnamefont
  {Chu}}, \bibinfo {author} {\bibfnamefont {J.}~\bibnamefont {Will}}, \bibinfo
  {author} {\bibfnamefont {J.}~\bibnamefont {Arlt}}, \bibinfo {author}
  {\bibfnamefont {C.}~\bibnamefont {Klempt}}, \ and\ \bibinfo {author}
  {\bibfnamefont {A.~M.}\ \bibnamefont {Rey}},\ }\href
  {https://journals.aps.org/prl/abstract/10.1103/PhysRevLett.125.240504}
  {\bibfield  {journal} {\bibinfo  {journal} {Physical Review Letters}\
  }\textbf {\bibinfo {volume} {125}},\ \bibinfo {pages} {240504} (\bibinfo
  {year} {2020})}\BibitemShut {NoStop}%
\bibitem [{\citenamefont {Muniz}\ \emph {et~al.}(2020)\citenamefont {Muniz},
  \citenamefont {Barberena}, \citenamefont {Lewis-Swan}, \citenamefont {Young},
  \citenamefont {Cline}, \citenamefont {Rey},\ and\ \citenamefont
  {Thompson}}]{dpt2}%
  \BibitemOpen
  \bibfield  {author} {\bibinfo {author} {\bibfnamefont {J.~A.}\ \bibnamefont
  {Muniz}}, \bibinfo {author} {\bibfnamefont {D.}~\bibnamefont {Barberena}},
  \bibinfo {author} {\bibfnamefont {R.~J.}\ \bibnamefont {Lewis-Swan}},
  \bibinfo {author} {\bibfnamefont {D.~J.}\ \bibnamefont {Young}}, \bibinfo
  {author} {\bibfnamefont {J.~R.~K.}\ \bibnamefont {Cline}}, \bibinfo {author}
  {\bibfnamefont {A.~M.}\ \bibnamefont {Rey}}, \ and\ \bibinfo {author}
  {\bibfnamefont {J.~K.}\ \bibnamefont {Thompson}},\ }\href
  {https://www.nature.com/articles/s41586-020-2224-x} {\bibfield  {journal}
  {\bibinfo  {journal} {Nature}\ }\textbf {\bibinfo {volume} {580}},\ \bibinfo
  {pages} {602} (\bibinfo {year} {2020})}\BibitemShut {NoStop}%
\bibitem [{\citenamefont {Ma}\ \emph {et~al.}(2011)\citenamefont {Ma},
  \citenamefont {Wang}, \citenamefont {Sun},\ and\ \citenamefont
  {Nori}}]{ma2011quantum}%
  \BibitemOpen
  \bibfield  {author} {\bibinfo {author} {\bibfnamefont {J.}~\bibnamefont
  {Ma}}, \bibinfo {author} {\bibfnamefont {X.}~\bibnamefont {Wang}}, \bibinfo
  {author} {\bibfnamefont {C.-P.}\ \bibnamefont {Sun}}, \ and\ \bibinfo
  {author} {\bibfnamefont {F.}~\bibnamefont {Nori}},\ }\href
  {https://www.sciencedirect.com/science/article/pii/S0370157311002201}
  {\bibfield  {journal} {\bibinfo  {journal} {Physics Reports}\ }\textbf
  {\bibinfo {volume} {509}},\ \bibinfo {pages} {89} (\bibinfo {year}
  {2011})}\BibitemShut {NoStop}%
\bibitem [{\citenamefont {Pezz{\`e}}\ \emph {et~al.}(2018)\citenamefont
  {Pezz{\`e}}, \citenamefont {Smerzi}, \citenamefont {Oberthaler},
  \citenamefont {Schmied},\ and\ \citenamefont {Treutlein}}]{pezze2018quantum}%
  \BibitemOpen
  \bibfield  {author} {\bibinfo {author} {\bibfnamefont {L.}~\bibnamefont
  {Pezz{\`e}}}, \bibinfo {author} {\bibfnamefont {A.}~\bibnamefont {Smerzi}},
  \bibinfo {author} {\bibfnamefont {M.~K.}\ \bibnamefont {Oberthaler}},
  \bibinfo {author} {\bibfnamefont {R.}~\bibnamefont {Schmied}}, \ and\
  \bibinfo {author} {\bibfnamefont {P.}~\bibnamefont {Treutlein}},\ }\href@noop
  {} {\bibfield  {journal} {\bibinfo  {journal} {Reviews of Modern Physics}\
  }\textbf {\bibinfo {volume} {90}},\ \bibinfo {pages} {035005} (\bibinfo
  {year} {2018})}\BibitemShut {NoStop}%
\bibitem [{\citenamefont {Li}\ \emph {et~al.}(2022)\citenamefont {Li},
  \citenamefont {Colombo}, \citenamefont {Shu}, \citenamefont {Velez},
  \citenamefont {Pilatowsky-Cameo}, \citenamefont {Schmied}, \citenamefont
  {Choi}, \citenamefont {Lukin}, \citenamefont {Pedrozo-Pe{\~n}afiel},\ and\
  \citenamefont {Vuleti{\'c}}}]{li2022improving}%
  \BibitemOpen
  \bibfield  {author} {\bibinfo {author} {\bibfnamefont {Z.}~\bibnamefont
  {Li}}, \bibinfo {author} {\bibfnamefont {S.}~\bibnamefont {Colombo}},
  \bibinfo {author} {\bibfnamefont {C.}~\bibnamefont {Shu}}, \bibinfo {author}
  {\bibfnamefont {G.}~\bibnamefont {Velez}}, \bibinfo {author} {\bibfnamefont
  {S.}~\bibnamefont {Pilatowsky-Cameo}}, \bibinfo {author} {\bibfnamefont
  {R.}~\bibnamefont {Schmied}}, \bibinfo {author} {\bibfnamefont
  {S.}~\bibnamefont {Choi}}, \bibinfo {author} {\bibfnamefont {M.}~\bibnamefont
  {Lukin}}, \bibinfo {author} {\bibfnamefont {E.}~\bibnamefont
  {Pedrozo-Pe{\~n}afiel}}, \ and\ \bibinfo {author} {\bibfnamefont
  {V.}~\bibnamefont {Vuleti{\'c}}},\ }\href {https://arxiv.org/abs/2212.13880}
  {\bibfield  {journal} {\bibinfo  {journal} {arXiv:2212.13880}\ } (\bibinfo
  {year} {2022})}\BibitemShut {NoStop}%
\bibitem [{\citenamefont {Gross}\ \emph {et~al.}(2011)\citenamefont {Gross},
  \citenamefont {Strobel}, \citenamefont {Nicklas}, \citenamefont {Zibold},
  \citenamefont {Bar-Gill}, \citenamefont {Kurizki},\ and\ \citenamefont
  {Oberthaler}}]{gross2011atomic}%
  \BibitemOpen
  \bibfield  {author} {\bibinfo {author} {\bibfnamefont {C.}~\bibnamefont
  {Gross}}, \bibinfo {author} {\bibfnamefont {H.}~\bibnamefont {Strobel}},
  \bibinfo {author} {\bibfnamefont {E.}~\bibnamefont {Nicklas}}, \bibinfo
  {author} {\bibfnamefont {T.}~\bibnamefont {Zibold}}, \bibinfo {author}
  {\bibfnamefont {N.}~\bibnamefont {Bar-Gill}}, \bibinfo {author}
  {\bibfnamefont {G.}~\bibnamefont {Kurizki}}, \ and\ \bibinfo {author}
  {\bibfnamefont {M.}~\bibnamefont {Oberthaler}},\ }\href
  {https://www.nature.com/articles/nature10654} {\bibfield  {journal} {\bibinfo
   {journal} {Nature}\ }\textbf {\bibinfo {volume} {480}},\ \bibinfo {pages}
  {219} (\bibinfo {year} {2011})}\BibitemShut {NoStop}%
\bibitem [{\citenamefont {L{\"u}cke}\ \emph {et~al.}(2011)\citenamefont
  {L{\"u}cke}, \citenamefont {Scherer}, \citenamefont {Kruse}, \citenamefont
  {Pezz{\'e}}, \citenamefont {Deuretzbacher}, \citenamefont {Hyllus},
  \citenamefont {Topic}, \citenamefont {Peise}, \citenamefont {Ertmer},
  \citenamefont {Arlt} \emph {et~al.}}]{lucke2011twin}%
  \BibitemOpen
  \bibfield  {author} {\bibinfo {author} {\bibfnamefont {B.}~\bibnamefont
  {L{\"u}cke}}, \bibinfo {author} {\bibfnamefont {M.}~\bibnamefont {Scherer}},
  \bibinfo {author} {\bibfnamefont {J.}~\bibnamefont {Kruse}}, \bibinfo
  {author} {\bibfnamefont {L.}~\bibnamefont {Pezz{\'e}}}, \bibinfo {author}
  {\bibfnamefont {F.}~\bibnamefont {Deuretzbacher}}, \bibinfo {author}
  {\bibfnamefont {P.}~\bibnamefont {Hyllus}}, \bibinfo {author} {\bibfnamefont
  {O.}~\bibnamefont {Topic}}, \bibinfo {author} {\bibfnamefont
  {J.}~\bibnamefont {Peise}}, \bibinfo {author} {\bibfnamefont
  {W.}~\bibnamefont {Ertmer}}, \bibinfo {author} {\bibfnamefont
  {J.}~\bibnamefont {Arlt}},  \emph {et~al.},\ }\href
  {https://www.science.org/doi/10.1126/science.1208798} {\bibfield  {journal}
  {\bibinfo  {journal} {Science}\ }\textbf {\bibinfo {volume} {334}},\ \bibinfo
  {pages} {773} (\bibinfo {year} {2011})}\BibitemShut {NoStop}%
\bibitem [{\citenamefont {Periwal}\ \emph {et~al.}(2021)\citenamefont
  {Periwal}, \citenamefont {Cooper}, \citenamefont {Kunkel}, \citenamefont
  {Wienand}, \citenamefont {Davis},\ and\ \citenamefont
  {Schleier-Smith}}]{periwal2021programmable}%
  \BibitemOpen
  \bibfield  {author} {\bibinfo {author} {\bibfnamefont {A.}~\bibnamefont
  {Periwal}}, \bibinfo {author} {\bibfnamefont {E.~S.}\ \bibnamefont {Cooper}},
  \bibinfo {author} {\bibfnamefont {P.}~\bibnamefont {Kunkel}}, \bibinfo
  {author} {\bibfnamefont {J.~F.}\ \bibnamefont {Wienand}}, \bibinfo {author}
  {\bibfnamefont {E.~J.}\ \bibnamefont {Davis}}, \ and\ \bibinfo {author}
  {\bibfnamefont {M.}~\bibnamefont {Schleier-Smith}},\ }\href
  {https://www.nature.com/articles/s41586-021-04156-0} {\bibfield  {journal}
  {\bibinfo  {journal} {Nature}\ }\textbf {\bibinfo {volume} {600}},\ \bibinfo
  {pages} {630} (\bibinfo {year} {2021})}\BibitemShut {NoStop}%
\bibitem [{\citenamefont {Finger}\ \emph {et~al.}(2023)\citenamefont {Finger},
  \citenamefont {Rosa-Medina}, \citenamefont {Reiter}, \citenamefont
  {Christodoulou}, \citenamefont {Donner},\ and\ \citenamefont
  {Esslinger}}]{finger2023spin}%
  \BibitemOpen
  \bibfield  {author} {\bibinfo {author} {\bibfnamefont {F.}~\bibnamefont
  {Finger}}, \bibinfo {author} {\bibfnamefont {R.}~\bibnamefont {Rosa-Medina}},
  \bibinfo {author} {\bibfnamefont {N.}~\bibnamefont {Reiter}}, \bibinfo
  {author} {\bibfnamefont {P.}~\bibnamefont {Christodoulou}}, \bibinfo {author}
  {\bibfnamefont {T.}~\bibnamefont {Donner}}, \ and\ \bibinfo {author}
  {\bibfnamefont {T.}~\bibnamefont {Esslinger}},\ }\href
  {https://arxiv.org/abs/2303.11326} {\bibfield  {journal} {\bibinfo  {journal}
  {arXiv:2303.11326}\ } (\bibinfo {year} {2023})}\BibitemShut {NoStop}%
\bibitem [{\citenamefont {Panda}\ \emph {et~al.}(2022)\citenamefont {Panda},
  \citenamefont {Tao}, \citenamefont {Egelhoff}, \citenamefont {Ceja},
  \citenamefont {Xu},\ and\ \citenamefont {M{\"u}ller}}]{panda2022quantum}%
  \BibitemOpen
  \bibfield  {author} {\bibinfo {author} {\bibfnamefont {C.~D.}\ \bibnamefont
  {Panda}}, \bibinfo {author} {\bibfnamefont {M.}~\bibnamefont {Tao}}, \bibinfo
  {author} {\bibfnamefont {J.}~\bibnamefont {Egelhoff}}, \bibinfo {author}
  {\bibfnamefont {M.}~\bibnamefont {Ceja}}, \bibinfo {author} {\bibfnamefont
  {V.}~\bibnamefont {Xu}}, \ and\ \bibinfo {author} {\bibfnamefont
  {H.}~\bibnamefont {M{\"u}ller}},\ }\href {https://arxiv.org/abs/2210.07289}
  {\bibfield  {journal} {\bibinfo  {journal} {arXiv:2210.07289}\ } (\bibinfo
  {year} {2022})}\BibitemShut {NoStop}%
\bibitem [{\citenamefont {Luo}\ \emph {et~al.}(2023)\citenamefont {Luo},
  \citenamefont {Zhang}, \citenamefont {Koh}, \citenamefont {Wilson},
  \citenamefont {Chu}, \citenamefont {Holland}, \citenamefont {Rey},\ and\
  \citenamefont {Thompson}}]{luo2023cavity}%
  \BibitemOpen
  \bibfield  {author} {\bibinfo {author} {\bibfnamefont {C.}~\bibnamefont
  {Luo}}, \bibinfo {author} {\bibfnamefont {H.}~\bibnamefont {Zhang}}, \bibinfo
  {author} {\bibfnamefont {V.~P.}\ \bibnamefont {Koh}}, \bibinfo {author}
  {\bibfnamefont {J.~D.}\ \bibnamefont {Wilson}}, \bibinfo {author}
  {\bibfnamefont {A.}~\bibnamefont {Chu}}, \bibinfo {author} {\bibfnamefont
  {M.~J.}\ \bibnamefont {Holland}}, \bibinfo {author} {\bibfnamefont {A.~M.}\
  \bibnamefont {Rey}}, \ and\ \bibinfo {author} {\bibfnamefont {J.~K.}\
  \bibnamefont {Thompson}},\ }\href {https://arxiv.org/abs/2304.01411}
  {\bibfield  {journal} {\bibinfo  {journal} {arXiv:2304.01411}\ } (\bibinfo
  {year} {2023})}\BibitemShut {NoStop}%
\bibitem [{Note1()}]{Note1}%
  \BibitemOpen
  \bibinfo {note} {See Supplemental Material at [URL will be inserted by
  publisher] for details of effective Hamiltonian derivation, dynamical phase
  transition, undepleted pump approximation and experimental implementation,
  includes Ref. \cite {bo9,dpt1,dpt2,dpt3,cox2016spatially}}\BibitemShut
  {NoStop}%
\bibitem [{\citenamefont {Gl{\"u}ck}\ \emph {et~al.}(2002)\citenamefont
  {Gl{\"u}ck}, \citenamefont {Kolovsky},\ and\ \citenamefont
  {Korsch}}]{gluck2002wannier}%
  \BibitemOpen
  \bibfield  {author} {\bibinfo {author} {\bibfnamefont {M.}~\bibnamefont
  {Gl{\"u}ck}}, \bibinfo {author} {\bibfnamefont {A.~R.}\ \bibnamefont
  {Kolovsky}}, \ and\ \bibinfo {author} {\bibfnamefont {H.~J.}\ \bibnamefont
  {Korsch}},\ }\href
  {https://www.sciencedirect.com/science/article/pii/S0370157302001424}
  {\bibfield  {journal} {\bibinfo  {journal} {Physics Reports}\ }\textbf
  {\bibinfo {volume} {366}},\ \bibinfo {pages} {103} (\bibinfo {year}
  {2002})}\BibitemShut {NoStop}%
\bibitem [{\citenamefont {Aeppli}\ \emph {et~al.}(2022)\citenamefont {Aeppli},
  \citenamefont {Chu}, \citenamefont {Bothwell}, \citenamefont {Kennedy},
  \citenamefont {Kedar}, \citenamefont {He}, \citenamefont {Rey},\ and\
  \citenamefont {Ye}}]{dpt4}%
  \BibitemOpen
  \bibfield  {author} {\bibinfo {author} {\bibfnamefont {A.}~\bibnamefont
  {Aeppli}}, \bibinfo {author} {\bibfnamefont {A.}~\bibnamefont {Chu}},
  \bibinfo {author} {\bibfnamefont {T.}~\bibnamefont {Bothwell}}, \bibinfo
  {author} {\bibfnamefont {C.~J.}\ \bibnamefont {Kennedy}}, \bibinfo {author}
  {\bibfnamefont {D.}~\bibnamefont {Kedar}}, \bibinfo {author} {\bibfnamefont
  {P.}~\bibnamefont {He}}, \bibinfo {author} {\bibfnamefont {A.~M.}\
  \bibnamefont {Rey}}, \ and\ \bibinfo {author} {\bibfnamefont
  {J.}~\bibnamefont {Ye}},\ }\href {https://doi.org/10.1126/sciadv.adc9242}
  {\bibfield  {journal} {\bibinfo  {journal} {Science Advances}\ }\textbf
  {\bibinfo {volume} {8}},\ \bibinfo {pages} {eadc9242} (\bibinfo {year}
  {2022})}\BibitemShut {NoStop}%
\bibitem [{\citenamefont {Mivehvar}\ \emph {et~al.}(2021)\citenamefont
  {Mivehvar}, \citenamefont {Piazza}, \citenamefont {Donner},\ and\
  \citenamefont {Ritsch}}]{Helmut2}%
  \BibitemOpen
  \bibfield  {author} {\bibinfo {author} {\bibfnamefont {F.}~\bibnamefont
  {Mivehvar}}, \bibinfo {author} {\bibfnamefont {F.}~\bibnamefont {Piazza}},
  \bibinfo {author} {\bibfnamefont {T.}~\bibnamefont {Donner}}, \ and\ \bibinfo
  {author} {\bibfnamefont {H.}~\bibnamefont {Ritsch}},\ }\href
  {https://www.tandfonline.com/doi/full/10.1080/00018732.2021.1969727}
  {\bibfield  {journal} {\bibinfo  {journal} {Advances in Physics}\ }\textbf
  {\bibinfo {volume} {70}},\ \bibinfo {pages} {1} (\bibinfo {year}
  {2021})}\BibitemShut {NoStop}%
\bibitem [{\citenamefont {Colella}\ \emph {et~al.}(2022)\citenamefont
  {Colella}, \citenamefont {Kosior}, \citenamefont {Mivehvar},\ and\
  \citenamefont {Ritsch}}]{Helmut3}%
  \BibitemOpen
  \bibfield  {author} {\bibinfo {author} {\bibfnamefont {E.}~\bibnamefont
  {Colella}}, \bibinfo {author} {\bibfnamefont {A.}~\bibnamefont {Kosior}},
  \bibinfo {author} {\bibfnamefont {F.}~\bibnamefont {Mivehvar}}, \ and\
  \bibinfo {author} {\bibfnamefont {H.}~\bibnamefont {Ritsch}},\ }\href
  {https://journals.aps.org/prl/abstract/10.1103/PhysRevLett.128.070603}
  {\bibfield  {journal} {\bibinfo  {journal} {Physical Review Letters}\
  }\textbf {\bibinfo {volume} {128}},\ \bibinfo {pages} {070603} (\bibinfo
  {year} {2022})}\BibitemShut {NoStop}%
\bibitem [{\citenamefont {Rosa-Medina}\ \emph {et~al.}(2022)\citenamefont
  {Rosa-Medina}, \citenamefont {Ferri}, \citenamefont {Finger}, \citenamefont
  {Dogra}, \citenamefont {Kroeger}, \citenamefont {Lin}, \citenamefont
  {Chitra}, \citenamefont {Donner},\ and\ \citenamefont
  {Esslinger}}]{rosa2022}%
  \BibitemOpen
  \bibfield  {author} {\bibinfo {author} {\bibfnamefont {R.}~\bibnamefont
  {Rosa-Medina}}, \bibinfo {author} {\bibfnamefont {F.}~\bibnamefont {Ferri}},
  \bibinfo {author} {\bibfnamefont {F.}~\bibnamefont {Finger}}, \bibinfo
  {author} {\bibfnamefont {N.}~\bibnamefont {Dogra}}, \bibinfo {author}
  {\bibfnamefont {K.}~\bibnamefont {Kroeger}}, \bibinfo {author} {\bibfnamefont
  {R.}~\bibnamefont {Lin}}, \bibinfo {author} {\bibfnamefont {R.}~\bibnamefont
  {Chitra}}, \bibinfo {author} {\bibfnamefont {T.}~\bibnamefont {Donner}}, \
  and\ \bibinfo {author} {\bibfnamefont {T.}~\bibnamefont {Esslinger}},\ }\href
  {https://journals.aps.org/prl/abstract/10.1103/PhysRevLett.128.143602}
  {\bibfield  {journal} {\bibinfo  {journal} {Physical Review Letters}\
  }\textbf {\bibinfo {volume} {128}},\ \bibinfo {pages} {143602} (\bibinfo
  {year} {2022})}\BibitemShut {NoStop}%
\bibitem [{\citenamefont {Cox}\ \emph {et~al.}(2016)\citenamefont {Cox},
  \citenamefont {Greve}, \citenamefont {Wu},\ and\ \citenamefont
  {Thompson}}]{cox2016spatially}%
  \BibitemOpen
  \bibfield  {author} {\bibinfo {author} {\bibfnamefont {K.~C.}\ \bibnamefont
  {Cox}}, \bibinfo {author} {\bibfnamefont {G.~P.}\ \bibnamefont {Greve}},
  \bibinfo {author} {\bibfnamefont {B.}~\bibnamefont {Wu}}, \ and\ \bibinfo
  {author} {\bibfnamefont {J.~K.}\ \bibnamefont {Thompson}},\ }\href
  {https://journals.aps.org/pra/abstract/10.1103/PhysRevA.94.061601} {\bibfield
   {journal} {\bibinfo  {journal} {Physical Review A}\ }\textbf {\bibinfo
  {volume} {94}},\ \bibinfo {pages} {061601} (\bibinfo {year}
  {2016})}\BibitemShut {NoStop}%
\end{thebibliography}%


\providecommand{\noopsort}[1]{}\providecommand{\singleletter}[1]{#1}%
\begin{thebibliography}{29}%
\makeatletter
\providecommand \@ifxundefined [1]{%
 \@ifx{#1\undefined}
}%
\providecommand \@ifnum [1]{%
 \ifnum #1\expandafter \@firstoftwo
 \else \expandafter \@secondoftwo
 \fi
}%
\providecommand \@ifx [1]{%
 \ifx #1\expandafter \@firstoftwo
 \else \expandafter \@secondoftwo
 \fi
}%
\providecommand \natexlab [1]{#1}%
\providecommand \enquote  [1]{``#1''}%
\providecommand \bibnamefont  [1]{#1}%
\providecommand \bibfnamefont [1]{#1}%
\providecommand \citenamefont [1]{#1}%
\providecommand \href@noop [0]{\@secondoftwo}%
\providecommand \href [0]{\begingroup \@sanitize@url \@href}%
\providecommand \@href[1]{\@@startlink{#1}\@@href}%
\providecommand \@@href[1]{\endgroup#1\@@endlink}%
\providecommand \@sanitize@url [0]{\catcode `\\12\catcode `\$12\catcode
  `\&12\catcode `\#12\catcode `\^12\catcode `\_12\catcode `\%12\relax}%
\providecommand \@@startlink[1]{}%
\providecommand \@@endlink[0]{}%
\providecommand \url  [0]{\begingroup\@sanitize@url \@url }%
\providecommand \@url [1]{\endgroup\@href {#1}{\urlprefix }}%
\providecommand \urlprefix  [0]{URL }%
\providecommand \Eprint [0]{\href }%
\providecommand \doibase [0]{https://doi.org/}%
\providecommand \selectlanguage [0]{\@gobble}%
\providecommand \bibinfo  [0]{\@secondoftwo}%
\providecommand \bibfield  [0]{\@secondoftwo}%
\providecommand \translation [1]{[#1]}%
\providecommand \BibitemOpen [0]{}%
\providecommand \bibitemStop [0]{}%
\providecommand \bibitemNoStop [0]{.\EOS\space}%
\providecommand \EOS [0]{\spacefactor3000\relax}%
\providecommand \BibitemShut  [1]{\csname bibitem#1\endcsname}%
\let\auto@bib@innerbib\@empty
%</preamble>
\bibitem [{\citenamefont {Chu}\ \emph {et~al.}(2020)\citenamefont {Chu},
  \citenamefont {Will}, \citenamefont {Arlt}, \citenamefont {Klempt},\ and\
  \citenamefont {Rey}}]{dpt1}%
  \BibitemOpen
  \bibfield  {author} {\bibinfo {author} {\bibfnamefont {A.}~\bibnamefont
  {Chu}}, \bibinfo {author} {\bibfnamefont {J.}~\bibnamefont {Will}}, \bibinfo
  {author} {\bibfnamefont {J.}~\bibnamefont {Arlt}}, \bibinfo {author}
  {\bibfnamefont {C.}~\bibnamefont {Klempt}},\ and\ \bibinfo {author}
  {\bibfnamefont {A.~M.}\ \bibnamefont {Rey}},\ }\bibfield  {title} {\bibinfo
  {title} {Simulation of xxz spin models using sideband transitions in trapped
  bosonic gases},\ }\href
  {https://journals.aps.org/prl/abstract/10.1103/PhysRevLett.125.240504}
  {\bibfield  {journal} {\bibinfo  {journal} {Physical Review Letters}\
  }\textbf {\bibinfo {volume} {125}},\ \bibinfo {pages} {240504} (\bibinfo
  {year} {2020})}\BibitemShut {NoStop}%
\bibitem [{\citenamefont {Muniz}\ \emph {et~al.}(2020)\citenamefont {Muniz},
  \citenamefont {Barberena}, \citenamefont {Lewis-Swan}, \citenamefont {Young},
  \citenamefont {Cline}, \citenamefont {Rey},\ and\ \citenamefont
  {Thompson}}]{dpt2}%
  \BibitemOpen
  \bibfield  {author} {\bibinfo {author} {\bibfnamefont {J.~A.}\ \bibnamefont
  {Muniz}}, \bibinfo {author} {\bibfnamefont {D.}~\bibnamefont {Barberena}},
  \bibinfo {author} {\bibfnamefont {R.~J.}\ \bibnamefont {Lewis-Swan}},
  \bibinfo {author} {\bibfnamefont {D.~J.}\ \bibnamefont {Young}}, \bibinfo
  {author} {\bibfnamefont {J.~R.~K.}\ \bibnamefont {Cline}}, \bibinfo {author}
  {\bibfnamefont {A.~M.}\ \bibnamefont {Rey}},\ and\ \bibinfo {author}
  {\bibfnamefont {J.~K.}\ \bibnamefont {Thompson}},\ }\bibfield  {title}
  {\bibinfo {title} {Exploring dynamical phase transitions with cold atoms in
  an optical cavity},\ }\href
  {https://www.nature.com/articles/s41586-020-2224-x} {\bibfield  {journal}
  {\bibinfo  {journal} {Nature}\ }\textbf {\bibinfo {volume} {580}},\ \bibinfo
  {pages} {602} (\bibinfo {year} {2020})}\BibitemShut {NoStop}%
\bibitem [{\citenamefont {Alberti}\ \emph {et~al.}(2010)\citenamefont
  {Alberti}, \citenamefont {Ferrari}, \citenamefont {Ivanov}, \citenamefont
  {Chiofalo},\ and\ \citenamefont {Tino}}]{bo9}%
  \BibitemOpen
  \bibfield  {author} {\bibinfo {author} {\bibfnamefont {A.}~\bibnamefont
  {Alberti}}, \bibinfo {author} {\bibfnamefont {G.}~\bibnamefont {Ferrari}},
  \bibinfo {author} {\bibfnamefont {V.~V.}\ \bibnamefont {Ivanov}}, \bibinfo
  {author} {\bibfnamefont {M.~L.}\ \bibnamefont {Chiofalo}},\ and\ \bibinfo
  {author} {\bibfnamefont {G.~M.}\ \bibnamefont {Tino}},\ }\bibfield  {title}
  {\bibinfo {title} {Atomic wave packets in amplitude-modulated vertical
  optical lattices},\ }\href
  {https://iopscience.iop.org/article/10.1088/1367-2630/12/6/065037} {\bibfield
   {journal} {\bibinfo  {journal} {New Journal of Physics}\ }\textbf {\bibinfo
  {volume} {12}},\ \bibinfo {pages} {065037} (\bibinfo {year}
  {2010})}\BibitemShut {NoStop}%
\bibitem [{\citenamefont {Chu}\ \emph {et~al.}(2021)\citenamefont {Chu},
  \citenamefont {He}, \citenamefont {Thompson},\ and\ \citenamefont
  {Rey}}]{dpt3}%
  \BibitemOpen
  \bibfield  {author} {\bibinfo {author} {\bibfnamefont {A.}~\bibnamefont
  {Chu}}, \bibinfo {author} {\bibfnamefont {P.}~\bibnamefont {He}}, \bibinfo
  {author} {\bibfnamefont {J.~K.}\ \bibnamefont {Thompson}},\ and\ \bibinfo
  {author} {\bibfnamefont {A.~M.}\ \bibnamefont {Rey}},\ }\bibfield  {title}
  {\bibinfo {title} {Quantum enhanced cavity qed interferometer with partially
  delocalized atoms in lattices},\ }\href
  {https://journals.aps.org/prl/abstract/10.1103/PhysRevLett.127.210401}
  {\bibfield  {journal} {\bibinfo  {journal} {Physical Review Letters}\
  }\textbf {\bibinfo {volume} {127}},\ \bibinfo {pages} {210401} (\bibinfo
  {year} {2021})}\BibitemShut {NoStop}%
\bibitem [{\citenamefont {Cox}\ \emph {et~al.}(2016)\citenamefont {Cox},
  \citenamefont {Greve}, \citenamefont {Wu},\ and\ \citenamefont
  {Thompson}}]{cox2016spatially}%
  \BibitemOpen
  \bibfield  {author} {\bibinfo {author} {\bibfnamefont {K.~C.}\ \bibnamefont
  {Cox}}, \bibinfo {author} {\bibfnamefont {G.~P.}\ \bibnamefont {Greve}},
  \bibinfo {author} {\bibfnamefont {B.}~\bibnamefont {Wu}},\ and\ \bibinfo
  {author} {\bibfnamefont {J.~K.}\ \bibnamefont {Thompson}},\ }\bibfield
  {title} {\bibinfo {title} {Spatially homogeneous entanglement for matter-wave
  interferometry created with time-averaged measurements},\ }\href
  {https://journals.aps.org/pra/abstract/10.1103/PhysRevA.94.061601} {\bibfield
   {journal} {\bibinfo  {journal} {Physical Review A}\ }\textbf {\bibinfo
  {volume} {94}},\ \bibinfo {pages} {061601} (\bibinfo {year}
  {2016})}\BibitemShut {NoStop}%
\end{thebibliography}%

\end{document}

\title{Control  and amplification of Bloch oscillations via photon-mediated interactions: Supplemental Materials}

\author{Haoqing Zhang}
\affiliation{JILA, NIST and Department of Physics, University of Colorado, Boulder, Colorado 80309, USA}
\affiliation{Center for Theory of Quantum Matter, University of Colorado, Boulder, Colorado 80309, USA}

\author{Anjun Chu}
\affiliation{JILA, NIST and Department of Physics, University of Colorado, Boulder, Colorado 80309, USA}
\affiliation{Center for Theory of Quantum Matter, University of Colorado, Boulder, Colorado 80309, USA}

\author{Chengyi Luo}
\affiliation{JILA, NIST and Department of Physics, University of Colorado, Boulder, Colorado 80309, USA}

\author{James K. Thompson}
\affiliation{JILA, NIST and Department of Physics, University of Colorado, Boulder, Colorado 80309, USA}

\author{Ana Maria Rey}
\affiliation{JILA, NIST and Department of Physics, University of Colorado, Boulder, Colorado 80309, USA}
\affiliation{Center for Theory of Quantum Matter, University of Colorado, Boulder, Colorado 80309, USA}

\date{\today}

\maketitle
\section{Cavity QED with Wannier-Stark state}

\subsection{Dispersive coupling between atoms and cavity}
In the main text, we considered $N$ ultracold atoms trapped in a standing-wave optical cavity along the vertical direction $\hat{z}$. The atoms are assumed to be confined in the ground band of the one-dimensional lattice with lattice depth $V_0$ and wave vector $k_l=2\pi/\lambda_l$. A single internal level $\left|g\right\rangle$ in the atomic ground manifold is coupled to an atomic excited state $|e\rangle$ with a transition energy $\hbar\omega_{0}=\hbar(\omega_{e}-\omega_{g})$, via a single cavity mode $\hat{a}$ with angular frequency $\omega_c$ and wavelength $\lambda_c$. 
The atom-cavity coupling has spatial dependence $\mathcal{G}(z)=\mathcal{G}_0 \sin(k_c z)$, with $k_c=2\pi/\lambda_c$. The cavity mode is coherently driven by an external light field with detuning $\Delta_c=\omega_p-\omega_c$ from the bare cavity mode, which generates a net injected field in the cavity with amplitude $\eta_p$. The full atom-cavity Hamiltonian is given as $H=\hat{H}_{{\mathrm{atom}}}+\hat{H}_{{\mathrm{light}}}+\hat{H}_{{\mathrm{int}}}$. each of the terms can be written as :
\begin{align}
\hat{H}_{{\mathrm{light}}} & =\hbar\left(\eta_{p}\hat{a}^{\dagger}e^{-i\omega_{p}t}+\eta_{p}^{*}\hat{a}e^{i\omega_{p}t}\right)+\hbar\omega_{c}\hat{a}^{\dagger}\hat{a}\\
\hat{H}_{{\mathrm{atom}}} & = \sum_{\tau=g,e} \int dz \ \hat{\psi}_{\tau}^{\dagger}(z)\left[\frac{p^{2}}{2M}+V(z)+\hbar\omega_{\tau}\right]\hat{\psi}_{\tau}(z)\\
\hat{H}_{{\mathrm{int}}} & =\hbar\int dz\ \mathcal{G}_0\sin k_{c}z\left[\hat{a}\hat{\psi}_{e}^{\dagger}(z)\hat{\psi}_{g}(z)+\hat{a}^{\dagger}\hat{\psi}_{g}^{\dagger}(z)\hat{\psi}_{e}(z)\right].
\end{align}
Here $V(z)=Mgz+V_{0}\sin^{2}k_{l}z$ describes the external potentials experienced by  the atoms. $\hat{\psi}_{e(g)}^{\dagger}(z)$ is
the field operator that  creates an atom in the state  $e (g)$ at position $z$, $\omega_{e(g)}$. Under the rotating frame of  the pump field (set by the Hamiltonian $H_{0}=\hbar\omega_{p}\hat{a}^{\dagger}\hat{a}+\hbar\omega_{p}\int dz\ \hat{\psi}_{e}^{\dagger}(z)\hat{\psi}_{e}(z)$), the system's Hamiltonian takes the following form:
\begin{align}
    \hat{H} &= \hbar\left(\eta_{p}\hat{a}^{\dagger}+\eta_{p}^{*}\hat{a}\right)-\Delta_{c}\hbar\hat{a}^{\dagger}\hat{a}-\hbar\Delta_{0}\int dz\ \hat{\psi}_{e}^{\dagger}(z)\hat{\psi}_{e}(z) + \sum_{\tau=g,e} \int dz \ \hat{\psi}_{\tau}^{\dagger}(z)\left[\frac{p^{2}}{2M}+V(z)\right]\hat{\psi}_{\tau}(z) \\
    &+ \hbar\int dz\mathcal{G}_0\sin k_{c}z\left[\hat{a} \hat{\psi}_{e}^{\dagger}(z) \hat{\psi}_{g}(z)+\hat{a}^{\dagger} \hat{\psi}_{g}^{\dagger}(z)\hat{\psi}_{e}(z)\right],
\end{align} where we   defined  the detuning of the pump from the atomic transition as  $\Delta_{0}=\omega_{p}-\omega_{0}$.

Furthermore, under the assumption $\Delta_{0}\gg\mathcal{G}_0\sqrt{\left\langle \hat{a}^{\dagger}\hat{a}\right\rangle }$ and $\Delta_{0}\gg\gamma$  with $\gamma$ the excited state spontaneous emission rate,
the excited state population remains   negligible during the relevant time scales. In this limit we can adiabatic eliminate the excited state $\left|e\right\rangle$  ($\hat{\psi}_e(z) \approx \mathcal{G}_0 \hat{a} \hat{\psi}_g(z) \sin k_c z / \Delta_0$), which leads to the following effective Hamiltonian acting on the ground state $\left|g\right\rangle$ manifold only,
\begin{equation}
\hat{H} = -\hbar\Delta_{c}\hat{a}^{\dagger}\hat{a}+\hbar\left(\eta_{p}\hat{a}^{\dagger}+\eta_{p}^{*}\hat{a}\right)+\int dz\ \hat{\psi}_{g}^{\dagger}(z)\left[\frac{\hbar(\mathcal{G}_0\sin k_{c}z)^{2}}{\Delta_{0}}\hat{a}^{\dagger}\hat{a} + \frac{p^{2}}{2M}+V(z)\right]\hat{\psi}_{g}(z). \label{eq:h1}
\end{equation}

 In  the tight-binding limit, the resulting  single-particle eigenstates of the Hamiltonian  $p^2/2m+V(z)$ become the so-called  Wannier-Stark (WS) states $\left|\phi_n\right\rangle$ ($n\in\mathbb{Z}$):
\begin{equation}
    E_n = M g a_l n,\quad \phi_n(z) = \sum_m \mathcal{J}_{m-n}\left(\frac{2J_0}{M g a_l}\right) w(z - m a_l). \label{eq:ws}
\end{equation}
Here $\mathcal{J}_n$ denotes  the Bessel function of the first kind, $J_0$ denotes the nearest-neighbor couplings in the ground band, $a_l=\lambda_l/2$ is  the lattice spacing and $w(z)$ is  the ground band Wannier function.  We will also use $E_R=(\hbar k_l)^2/2M$ for the atomic recoil energy. 
\begin{figure}
	\centering
	\includegraphics[width=0.5\columnwidth]{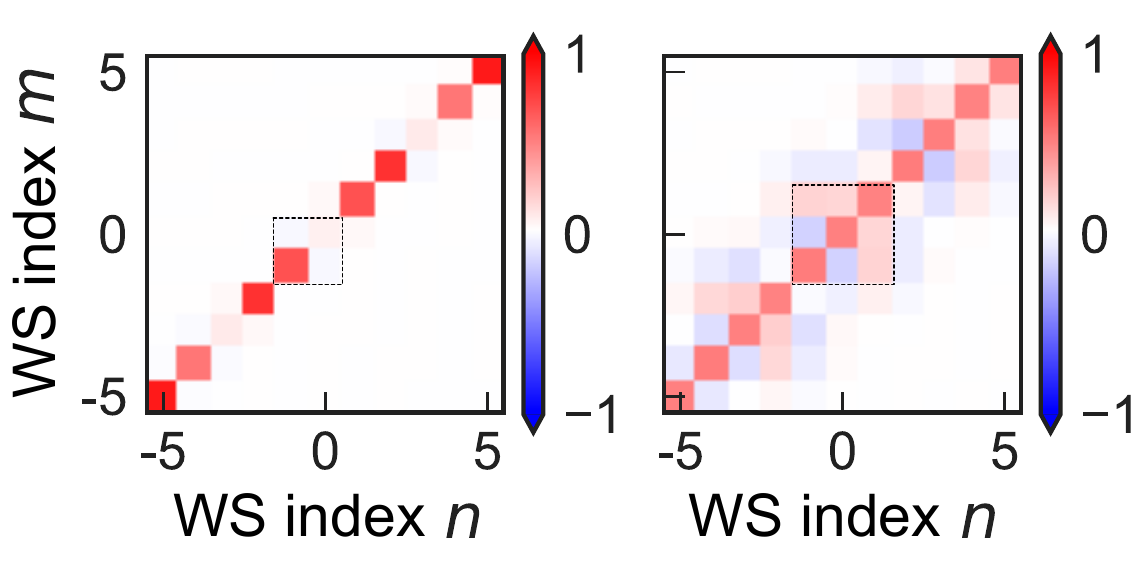}
	\caption{\label{fig:Jmn} The coupling coefficient $J_{m,n}$ for $^{87}\mathrm{Rb}$ atoms ($\lambda_l=532$ nm, $\lambda_c=780$ nm). Left: $V_0=20E_R$ and right:  $V_0=6E_R$. Start from $\left|\phi_0\right\rangle$, the many-body dynamics mainly happens within the dashed square for either two-level model (left, deep lattice region for dynamical phase transitions) and three-level model (right, shallow lattice region for amplification of Bloch oscillations).
	}
\end{figure} 

The field operator, when written in the WS basis takes the form, $\hat{\psi}_{g}(z)=\sum_{n}\hat{c}_{n}\phi_{n}(z)$, where $\hat{c}_{n}$ annihilates an atom in the state $\left|\phi_n\right\rangle$. In this basis we can  rewrite the Hamiltonian [Eq. \eqref{eq:h1}] as:
\begin{equation}
\hat{H}/\hbar=-\Delta_{c}\hat{a}^{\dagger}\hat{a}+\eta_{p}\hat{a}^{\dagger}+\eta_{p}^{*}\hat{a}+\frac{\mathcal{G}_0^{2}}{\Delta_{0}}\hat{a}^{\dagger}\hat{a}\sum_{m,n}J_{m,n}\hat{c}_{m}^{\dagger}\hat{c}_{n}+\omega_{B}\sum_{n}n\hat{c}_{n}^{\dagger}\hat{c}_{n}, \label{eq:h2}
\end{equation}
where  $J_{m,n}=\int dz\phi_{m}(z)\phi_{n}(z)\sin^{2}k_{c}z$ describes the overlap between the WS states $\left|\phi_m\right\rangle$, $\left|\phi_n\right\rangle$
weighted by the cavity field mode function. In Fig. \ref{fig:Jmn}  we show the value of these couplings  for the typical lattice depths we work in this paper. We define the effective particle number:
\begin{equation}
    \hat{N}_{\mathrm{eff}}=\sum_{m,n}J_{m,n}\hat{c}_{m}^{\dagger}\hat{c}_{n}, \label{eq:Neff}
\end{equation}
as the effective number of atoms coupled to the cavity, which shifts the cavity resonance frequency by $\mathcal{G}_0^2 \left\langle\hat{N}_{\mathrm{eff}}\right\rangle / \Delta_0$.

\subsection{Adiabatic elimination of cavity field}
Here we study the dynamics via Heisenberg
equations of motion using a Markovian approximation. We adiabatic eliminate the cavity field using the fact that  $\Delta_c$ sets the largest frequency scale and derive the effective atom-only Hamiltonian. To do that, we formally integrate the Heisenberg equation of motion of  the cavity mode operator $\hat{a}$ and photon number operators $\hat{a}^\dagger \hat{a}$, then plug them back into the Hamiltonian [Eq. \eqref{eq:h2}]. 
We  remove the fast rotating terms which  relax  much faster than the time it takes an atom to perform a BO. 

The Heisenberg-Langevin equation of the motion for the cavity mode $\hat{a}$ is given by:
\begin{equation}
    \frac{d}{dt}\hat{a}=i[\hat{H}/\hbar,\hat{a}]+(\frac{\kappa}{2}\hat{a}^{\dagger}+\hat{f}^{\dagger})[\hat{a},\hat{a}]-[\hat{a},\hat{a}^{\dagger}](\frac{\kappa}{2}\hat{a}+\hat{f})=i(\Delta_{c}-\frac{\mathcal{G}_{0}^{2}\hat{N}_{\mathrm{eff}}}{\Delta_{0}}) \hat{a} -i\eta_{p}-\frac{\kappa}{2}\hat{a} + \hat{f},
\end{equation}
with $\kappa$ for the cavity decay rate.
The above  equation captures  the dissipative dynamics  generated by $\kappa$  along with the quantum Langevin noise operator $\hat{f}$, which gives the formal solution for the cavity field operator:
\begin{equation}
\begin{aligned}
        \hat{a}&=-i\eta_{p}\exp\left[i\int_{0}^{t}d\tau\left(\hat{\Delta}+i\kappa/2\right)\right]\int_{0}^{t}dt^{\prime}\exp\left[-i\int_{0}^{t^{\prime}}d\tau\left(\hat{\Delta}+i\kappa/2\right)\right] + \hat{f}^\prime \\
        &\approx \frac{\eta_p}{\hat{\Delta}+i\kappa/2} + \hat{f}^\prime
\end{aligned}
\end{equation}
with
\begin{equation}
    \hat{\Delta}=\Delta_c -\frac{\mathcal{G}_{0}^{2}\hat{N}_{\mathrm{eff}}}{ \Delta_{0}},
\end{equation}
Here  $\hat{f}^\prime$ is  another quantum  Langevin noise operator. Below we consider the regime  $\Delta_c,\kappa \gg \omega_B$ where  the cavity-field dynamics evolve much faster than the atomic dynamics, thus it follows the latter adiabatically. As a result, we can obtain the formal solution for the cavity photon number operator:
\begin{equation}
    \hat{a}^\dagger \hat{a} = \frac{|\eta_p|^2}{\hat{\Delta}^2+(\kappa/2)^2} + \hat{g} \approx \frac{|\eta_p|^2}{\hat{\Delta}^2} .
\end{equation}
One more time  $\hat{g}$ represents a different quantum  Langevin noise operator. For the last approximation above, we focus of the regime  $\Delta_c\gg \kappa$ where  the unitary dynamics dominates  and we can ignore the dissipation process to leading order.

If we insert the above solution of the cavity field into Eq. \eqref{eq:h2}, the effective atom-only Hamiltonian in the Schrodinger picture can be written as:
\begin{equation}
\begin{aligned}
\hat{H}_{\mathrm{eff}} / \hbar & =\omega_B \sum_n n \hat{c}_n^{\dagger} \hat{c}_n+\eta_p \frac{\eta_p^*}{\hat{\Delta}}+\eta_p^* \frac{\eta_p}{\hat{\Delta}}+\left(\mathcal{G}_0^2 \hat{N}_{\mathrm{eff}} / \Delta_0-\Delta_c\right) \frac{\left|\eta_p\right|^2}{\hat{\Delta}^2} \\
& =\omega_B \sum_n n \hat{c}_n^{\dagger} \hat{c}_n+2 \frac{\left|\eta_p\right|^2}{\hat{\Delta}}-\hat{\Delta} \frac{\left|\eta_p\right|^2}{\hat{\Delta}^2} \\
& =\omega_B \sum_n n \hat{c}_n^{\dagger} \hat{c}_n+\frac{\left|\eta_p\right|^2}{\Delta_c-\mathcal{G}_0^2 \hat{N}_{\mathrm{eff}} / \Delta_0} \\
& \equiv \omega_B \sum_n n \hat{c}_n^{\dagger} \hat{c}_n+\hat{V}_{\mathrm{cav}}(\hat{N}_{\mathrm{eff}})\label{eq:heff}
\end{aligned}
\end{equation}
where $\hat{V}_\mathrm{cav}(\hat{N}_{\mathrm{eff}})=-( V N / \beta) / (1+\beta \hat{N}_{\mathrm{eff}}/N)$ is the dynamical potential induced by the cavity which depends on  the atomic motion. $\hat{V}_\mathrm{cav}$ is parameterized by the maximum AC Stark shift introduced by the bare cavity mode, $V=\mathcal{G}^2_0|\eta_p|^2/(\Delta_c^2\Delta_0)$, as well as by  the ratio between the maximum cavity shift and the bare cavity detuning, $\beta=-N\mathcal{G}_0^2/(\Delta_0\Delta_c)$.
We assume $\beta>0$ ($\Delta_0$ and $\Delta_c$ have opposite signs) to avoid hitting a resonance.

\begin{figure}
	\centering
	\includegraphics[width=0.8\columnwidth]{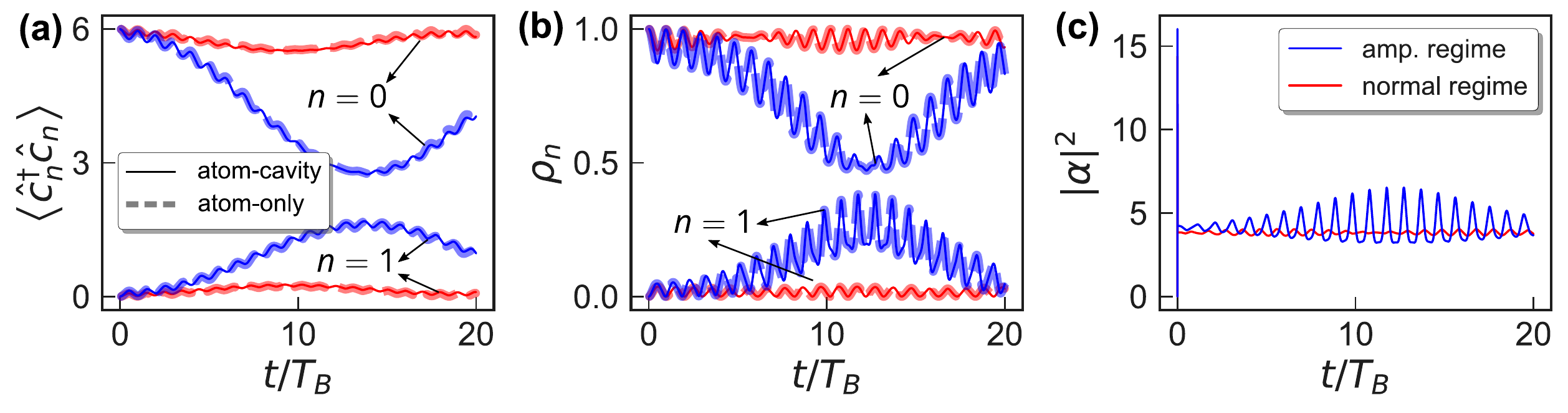}
	\caption{\label{fig:test} Benchmarks of the atom-cavity Hamiltonian [Eq.~\eqref{eq:h2}] with the effective atom-only Hamiltonian [Eq.~\eqref{eq:heff}]. The red curves for $V_0=5.8E_R$ (normal regime) and the blue curves for $V_0=6.2E_R$ (amplification regime) are used for all the simulations in the figure. 
    (a) Exact Diagonalization (ED) simulation of the dynamics for 6 particles in 3 WS states with the initial state $(\hat{c}^{\dagger}_0)^6 \left|\mathrm{vac}\right\rangle$. Populations $\left\langle\hat{c}_0^\dagger \hat{c}_0 \right\rangle$ (start from 6) as well as $\left\langle\hat{c}_1^\dagger \hat{c}_1\right\rangle$ (start from 0) for these two lattice depths are plotted. The solid lines are the exact simulations under the Hamiltonian Eq.~\eqref{eq:h2} (with the photon space $n_{\mathrm{cut}}=10$) and the dashed lines under the Hamiltonian Eq.~\eqref{eq:heff}. 
    (b) Mean-field dynamics of $\rho_n$ with initial state $|\phi_0\rangle$. The solid lines are simulated with the atom-cavity mean-field equations of motion [Eq.~\eqref{eq:eomac}] and the dashed lines are simulated with atom-only equations of motion [Eq.~\eqref{eq:eom2}]. Populations $\rho_0$ (start from 1) as well as $\rho_1$ (start from 0) for these two parameters are plotted.
    The differences between the atom-cavity and atom-only simulations can be ignored for both (a) and (b).
    (c) Mean-field evolution for the cavity photon number with the same parameters for the red and blue curves as in (b).
	}
\end{figure}

As a benchmark for the effective atom-only Hamiltonian derived in Eq.~\eqref{eq:heff}, we compare the exact dynamics for 6 particles in 3 WS states under Eq.~\eqref{eq:h2} and Eq.~\eqref{eq:heff} in  Fig.~\ref{fig:test}(a). In the simulation, we choose $\Delta_c=400\omega_B$, $\kappa=20\Delta_c$, $\eta_p=\Delta_c/10$ as well as $\mathcal{G}^2_0/\Delta_0=-100\omega_B$ in the atom-cavity simulation (dashed lines), which corresponds to $V=\mathcal{G}^2_0|\eta_p|^2/(\Delta_c^2\Delta_0)=\omega_B$ and $\beta=-N\mathcal{G}_0^2/(\Delta_0\Delta_c)=1.5$ in the atom-only simulation (solid lines). The simulation results match well with each other for the lattice depth in the normal regime (red curves) and amplification regime (blue curves), which verify the effectiveness of the atom-only Hamitlonian.

% \begin{equation}
%       \frac{d}{d t} \langle \hat{a} \rangle =i \langle[\hat{H}, \hat{a}]\rangle + \kappa ( \langle\hat{a}^\dagger \hat{a} \hat{a} \rangle - \frac{1}{2}  \langle\{\hat{a}^{\dagger}\hat{a},\hat{a}\}\rangle) =i \left(\Delta_c + i\kappa /2 - \frac{\mathcal{G}_{0}^{2}\hat{N}_{\mathrm{eff}}}{ \Delta_{0}}\right) \hat{a}-i \eta_{p} ,
% \end{equation}
% which gives the formal solution:
% \begin{equation}
%       \hat{a}(t)=\hat{a}(0) \exp \left[i \int_{0}^{t} d \tau (\hat{\Delta}(\tau)+ i\kappa/2)\right]-i \eta_{p} \exp \left[i \int_{0}^{t} d \tau (\hat{\Delta}(\tau)+ i\kappa/2)\right] \int_{0}^{t} d t^{\prime} \exp \left[-i \int_{0}^{t^{\prime}} d \tau (\hat{\Delta}(\tau)+ i\kappa/2)\right], \label{eq:at}
% \end{equation}
% with
% \begin{equation}
% \hat{\Delta}(t)=\Delta_c -\frac{\mathcal{G}_{0}^{2}\hat{N}_{\mathrm{eff}}(t)}{ \Delta_{0}}.
% \end{equation}

% We assume $\hat{a}(0)=0$ as the initial state so the first term in Eq. \eqref{eq:at} varnishes. Moreover,  $\hat{N}_{\mathrm{eff}}$ varies slowly compared with $\Delta_c$ and we can roughly assume it to be a constant during the time scale set by $\Delta_c$. In this way, we simply ignore the timed-order of $\hat{N}_{\mathrm{eff}}$ and treat it  as a c-number. The solution then becomes the following.
% \begin{equation}
%     \hat{a}(t)=\frac{\eta_p}{\hat{\Delta}+i\kappa/2}[1 - \exp(i\hat{\Delta}t - \kappa t/2)]. \label{eq:ass}
% \end{equation}
\subsection{Equations of motion for atoms}
To simulate the dynamics under Eq. \eqref{eq:heff}, we can calculate the equations of motion for the field operators $\hat{c}_m$ as:
\begin{equation}
    \begin{aligned}
    i \dot{\hat{c}}_{m} &=m \omega_{B} \hat{c}_{m} 
    - \frac{V N}{\beta} \left[\hat{c}_{m}, \frac{1}{\left(1+\beta \hat{N}_{\mathrm{eff}} / N\right)}\right] \\
    &=m \omega_{B} \hat{c}_{m}- \frac{V N}{\beta} \left[\hat{c}_{m}, 1 - \frac{\beta}{N} \hat{N}_{\mathrm{eff}}
    + (\frac{\beta}{N})^2 \hat{N}_{\mathrm{eff}}^{2} - (\frac{\beta}{N})^{3} \hat{N}_{\mathrm{eff}}^{3}+\cdots\right], \label{eq:eom1}
\end{aligned}
\end{equation}
Then we can simplify  the equations above with $\left[\hat{c}_{m}, \sum_{p, q} J_{p, q} \hat{c}_{p}^{\dagger} \hat{c}_{q}\right]=\sum_{n} J_{m, n} \hat{c}_{n}$:
\begin{equation}
    \begin{aligned}
    i \dot{\hat{c}}_{m} &\left.=m \omega_{B} \hat{c}_{m}
    - \frac{V N}{\beta}\left\{  -\frac{\beta}{N} \left [\hat{c}_{m}, \hat{N}_{\mathrm{eff}}\right] 
    + (\frac{\beta}{N})^2\left[\hat{c}_{m}, \hat{N}_{\mathrm{eff}}^{2}\right] 
    - (\frac{\beta}{N})^{3}\left[\hat{c}_{m}, \hat{N}_{\mathrm{eff}}^{3}\right]+\cdots\right]\right\} \\
    &\left.=m \omega_{B} \hat{c}_{m}
    - \frac{V N}{\beta} \left\{ - \frac{\beta}{N} \sum_n J_{m n} \hat{c}_{n}
    + (\frac{\beta}{N})^2 (\sum_n J_{m n} \hat{c}_{n}\hat{N}_{\mathrm{eff}}+\hat{N}_{\mathrm{eff}} \sum_n J_{m n} \hat{c}_{n})+\cdots\right]\right\}.
\end{aligned}
\end{equation}
Finally, we  apply the mean-field approximation to  the operators $\left\langle\sum_{n} J_{m n} \hat{c}_{n} \hat{N}_{\mathrm{eff}}\right\rangle\approx\left\langle\sum_{n} J_{m n} \hat{c}_{n}\right\rangle\left\langle\hat{N}_{\mathrm{eff}}\right\rangle$, and obtain  the  following equations of motion:
\begin{equation}
    i\left\langle\dot{\hat{c} }_{m}\right\rangle = m \omega_{B}  \left\langle\hat{c}_{m} \right\rangle
    + \frac{V}{\left(1+\beta\left\langle\hat{N}_{\mathrm{eff}}\right\rangle / N\right)^{2}} \sum_{m,n} J_{m,n} \left\langle\hat{c}_{n}\right\rangle \label{eq:eom2},
\end{equation}
All the results in the main text were  obtained by  solving the mean-field equations of motion written above. 

Meanwhile, the mean-field equations for the atom-cavity Hamiltonian [Eq.~\eqref{eq:h2}] is given by,
\begin{equation}
    \begin{aligned}
        i\dot{\alpha} &= - \left(\Delta_c + i\frac{\kappa}{2} - \frac{\mathcal{G}_0^2\left\langle\hat{N}_{\mathrm{eff}}\right\rangle}{\Delta_0}\right)\alpha + \eta_p \\
        i\left\langle\dot{\hat{c} }_{m}\right\rangle &= m \omega_{B}  \left\langle\hat{c}_{m} \right\rangle +\frac{\mathcal{G}_0^2}{\Delta_0}|\alpha|^2 \sum_{m,n} J_{m,n} \left\langle\hat{c}_{n}\right\rangle, \label{eq:eomac}
    \end{aligned}
\end{equation}
with $\alpha=\left\langle\hat{a}\right\rangle.$ We compare the mean-field dynamics Eq.~\eqref{eq:eom2} and Eq.~\eqref{eq:eomac} in Fig.~\ref{fig:test}(b). In the simulation, we choose reasonable experimental parameters $N=2\times 10^4$ atoms, $\Delta_c=2\pi\times 2\ $MHz, $\kappa=\Delta_c/20$, $\eta_p=3\Delta_c$ as well as $\mathcal{G}^2_0/\Delta_0=-2\pi\times 100\ $Hz in the atom-cavity simulation [Eq.~\eqref{eq:eomac}], which corresponds to $V=\mathcal{G}^2_0|\eta_p|^2/(\Delta_c^2\Delta_0)=1.57\omega_B$ and $\beta=-N\mathcal{G}_0^2/(\Delta_0\Delta_c)=1$ in the atom-only simulation [Eq.~\eqref{eq:eom2}]. Still, the simulations for both normal regime and amplification regime match with each other pretty well with the difference can be ignored, which again verifies the validation of the effective atom-only Hamiltonian. In Fig.~\ref{fig:test}(c), we plot the evolution of cavity photon number $|\alpha|^2$ which follows the atomic motion adiabatically.

\section{Dynamical phase transition with Wannier-Stark states}
In this part, we consider the deep lattice regime and discuss how to map the atom-only Hamiltonian to a spin model. As discussed in the main text, for a deep lattice $V_0=20E_R$, the WS states approach the Wannier orbitals which are localized. The overlap integral $J_{m,n}$ for $V_0=20E_R$ is shown in  the left panel  of Fig. \ref{fig:Jmn} with $J_{0,0}\approx0\ll J_{1,1} \approx J_{-1,-1}$ and $J_{1,0}\approx -J_{0,-1} \approx 0$. As a result non-trivial dynamics happens only for $V(J_{n,n}-J_{0,0}) + n\omega_B\approx 0$  when starting  from $\left|\phi_0\right\rangle$. Here we consider $V>0$ and deal with two bosonic modes $\hat{c}_{0}$, $\hat{c}_{-1}$. 
For simplicity, we define $\Omega_n=J_{n,n+1}$ as well as $\Delta_{n}=(J_{n,n}-J_{0,0})/2$. The spin operators are defined as follows,
\begin{equation}
\begin{aligned}
\hat{S}_{x}&=\frac{1}{2}(\hat{c}_{-1}^{\dagger}\hat{c}_{0}+\hat{c}_{0}^{\dagger}\hat{c}_{-1})\\
\hat{S}_{y}&=-\frac{i}{2}(\hat{c}_{-1}^{\dagger}\hat{c}_{0}-\hat{c}_{0}^{\dagger}\hat{c}_{-1})
\\\hat{S}_{z}&=\frac{1}{2}(\hat{c}_{-1}^{\dagger}\hat{c}_{-1}-\hat{c}_{0}^{\dagger}\hat{c}_{0}),
\end{aligned}    
\end{equation}
and  the total particle number $\hat{N}=\hat{c}_{-1}^{\dagger}\hat{c}_{-1}+\hat{c}_{0}^{\dagger}\hat{c}_{0}$. Such pseudospin operators satisfy the SU(2) algebra and we can rewrite the effective number operator as ($\bar{\omega}=(J_{-1,-1}+J_{0,0})/2$):
\begin{equation}
    \hat{N}_{\mathrm{eff}}=2\Omega_{-1} \hat{S}_{x} +2\Delta_{-1} \hat{S}_{z}+\bar{\omega}N,
\end{equation}
as well as the effective spin model from Eq. \eqref{eq:h2} in terms  only of $\hat{c}_{-1}$ and $\hat{c}_{0}$:
\begin{equation}
    \hat{H}_{\mathrm{eff}}/\hbar=-\omega_B \hat{S}_{z} + \hat{V}_\mathrm{cav}(\hat{N}_{\mathrm{eff}}) \label{eq:hspin}
\end{equation} 

Similar in Eq. \eqref{eq:eom1}, we can derive Heisenberg equations of motion for the collective spin operator $\hat{S}_{x,y,z}$. Using the  mean-field approximation which neglects the quantum correlation between different spins we  obtain,
\begin{equation}
\begin{aligned}
     \left\langle \dot{\hat{S}}_{x}\right\rangle &=(\omega_{B}-2\Delta_{-1} \tilde{V})\left\langle \hat{S}_{y}\right\rangle \\
    \left\langle \dot{\hat{S}}_{y}\right\rangle &=(2\Delta_{-1}   \tilde{V}-\omega_{B})\left\langle \hat{S}_{x}\right\rangle -2\Omega_{-1}     \tilde{V}\left\langle \hat{S}_{z}\right\rangle \\
    \left\langle \dot{\hat{S}}_{z}\right\rangle &=2\Omega_{-1}     \tilde{V}\left\langle \hat{S}_{y}\right\rangle.
\end{aligned}
\end{equation}

Moreover, if  we introduce the mean-field real variable $s_{\alpha}=2\left\langle \hat{S}_{\alpha}\right\rangle /N,\alpha\in\{x,y,z\}$, the above equations become:
\begin{equation}
\begin{aligned}
\dot{s}_{x}	&=(\omega_{B}-2\Delta_{-1}  \tilde{V} )s_{y} \\ 
\dot{s}_{y}	&=(2\Delta_{-1} \tilde{V} -\omega_{B})s_{x}-2\Omega_{-1} \tilde{V} s_{z} \\
\dot{s}_{z}	&=2\Omega_{-1}  \tilde{V} s_{y}, \label{eq:eom3}
\end{aligned}
\end{equation}
with 
\begin{equation}
    \tilde{V} = \frac{V}{\left(1+\beta N_{\mathrm{eff}}(t) / N\right)^{2}},
\end{equation}
here $N_{\mathrm{eff}}(t)=N(\Omega_{-1} s_{x}+\Delta_{-1} s_{z}+\bar{\omega})$. Later we will use the symbol $n_\mathrm{eff}\equiv\Omega_{-1} s_{x}+\Delta_{-1} s_{z}+\bar{\omega}$ for convenience. We compare the results from  Eq. \eqref{eq:eom2} and Eq. \eqref{eq:eom3} to numerical simulations of the full Hamiltonian and they match with each other, which means the two-mode approximation works in this case. 
Now, we discuss the dynamical phase transition predicted by Eq. \eqref{eq:eom3}.
Using both energy conservation as well as the identity  $(\hat{S}^x)^2+(\hat{S}^y)^2+(\hat{S}^z)^2=(\frac{N}{2}+1) \frac{N}{2}$ in the large $N$ limit, the real variable $(s_{x},s_{y},s_{z})$ with initial condition $s_{z}=-1,s_{x}=s_{y}=0$ satisfy the following two conservation laws:
\begin{align}
    s_{x}^{2}+s_{y}^{2}+s_{z}^{2}&=1 \\
    -\omega_B s_z - \frac{2 V/ \beta}{1+\beta n_\mathrm{eff}} 
    &=  \omega_B - \frac{2 V/ \beta}{1+\beta (\bar{\omega}-\Delta_{-1})}, \label{eq:con}
\end{align}
then we can express $s_{x,y,z}$ all as a function of $n_{\mathrm{eff}}$: 
\begin{equation}
    \begin{aligned}
    s_z(n_\mathrm{eff}) &= \frac{F(\bar{\omega} - \Delta_{-1}) - F(n_\mathrm{eff})}{\omega_B} - 1  \\
    s_x(n_\mathrm{eff}) &= \frac{n_\mathrm{eff} - \Delta_{-1} s_z(n_\mathrm{eff}) - \bar{\omega}}{\Omega_{-1}}  \\
    s_y^2(n_\mathrm{eff}) &= 1 - s_x^2(n_\mathrm{eff}) - s_z^2(n_\mathrm{eff}),
    \end{aligned}
\end{equation}
also we define a function,
\begin{equation}
    F(x) = \frac{2V/\beta}{1+\beta x}.
\end{equation}

\begin{figure}
	\centering	\includegraphics[width=0.4\columnwidth]{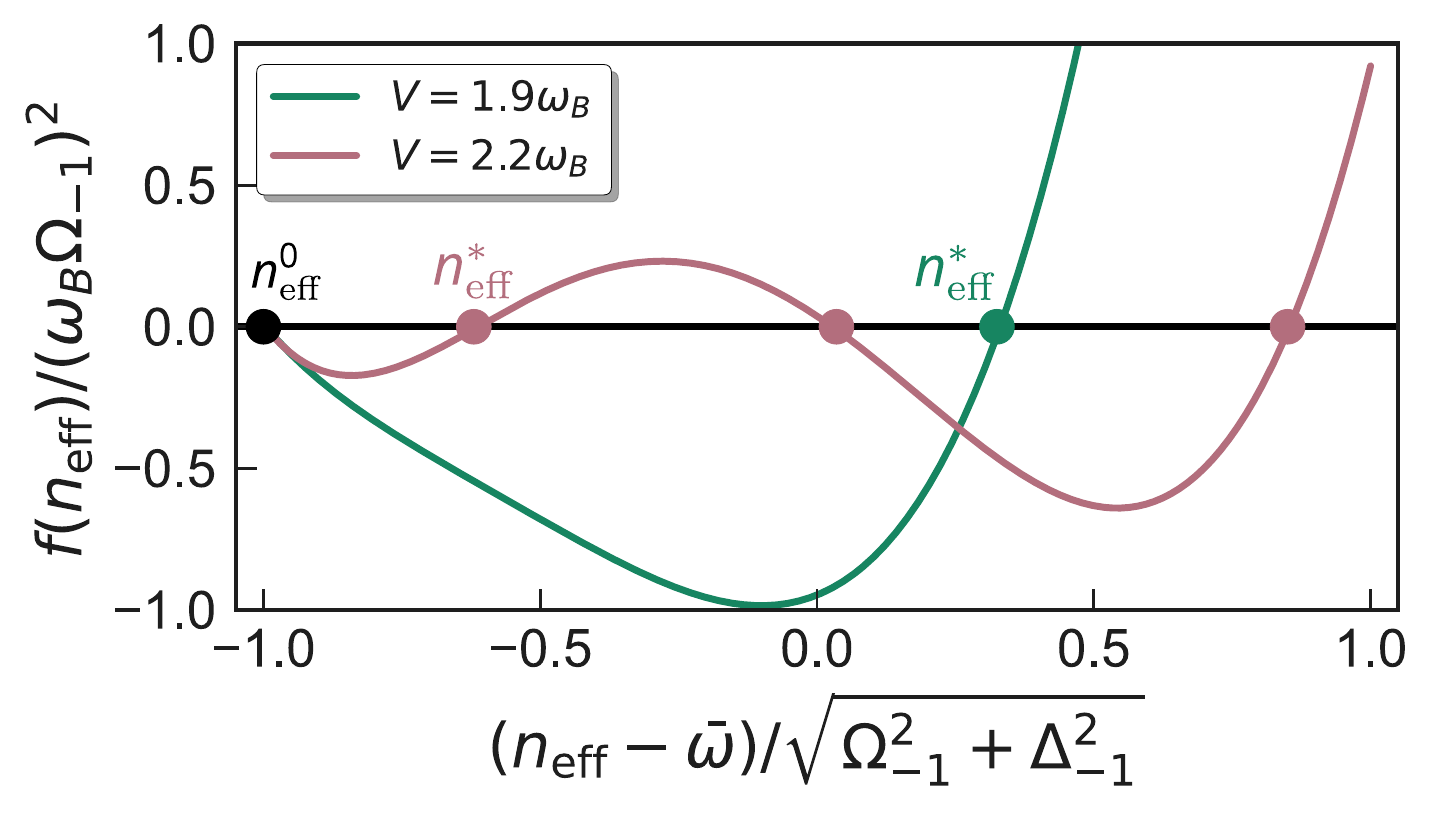}
	\caption{\label{fig:root}Number of roots for the effective potential $f(n_\mathrm{eff})$. In the case of $V=1.9\omega_B$, $\beta=0.5$ (green), $f(n_\mathrm{eff})$ has two roots; In the case of $V=2.1\omega_B$, $\beta=0.5$ (red), $f(n_\mathrm{eff})$ has four roots. 
The nearest root is labelled by $n_\mathrm{eff}^*$ and the jump of $n_\mathrm{eff}^*$ indicate DPTs.}
\end{figure}

The dynamics correspond to a classical particle moving in the external potential from Eq.~\eqref{eq:eom3}:
\begin{equation}
    (\dot{n_\mathrm{eff}})^2 + f(n_\mathrm{eff}) = 0, \label{eq:eomo}
\end{equation}
with the potential $f(n_\mathrm{eff})=- (\omega_B \Omega_{-1})^2 s_y^2(n_\mathrm{eff})$. 
The condition $f(n_\mathrm{eff})=0$ determines the roots and we find $n_\mathrm{eff}^0=\bar{\omega}-\Delta_{-1}$ is one of such root.
The effective potential can have either two or four solutions within the region $n_\mathrm{eff}\in[\bar{\omega}-\sqrt{\Omega^2_{-1}-\Delta^2_{-1}},\bar{\omega}+\sqrt{\Omega^2_{-1}+\Delta^2_{-1}}]$ shown in Fig.~\ref{fig:root}, and the dynamics of $n_\mathrm{neff}$ can be understood as the oscillations between $n_\mathrm{eff}^0$ and the nearest root $n_{\mathrm{eff}}^*$. Begin with a function $f(n_\mathrm{eff})$ with two roots, and continuously tune the parameters of $f(n_\mathrm{eff})$ so that two new roots appear in between, then a jump of the nearest root $n_\mathrm{eff}^*$ should occur during this process.
The dynamical paramagnetic phase corresponds to four roots, while the dynamical ferromagnetic phase corresponds to two roots. Deep in the dynamical paramagnetic phase, one can access the whole Bloch sphere ($n_\mathrm{neff}^*\approx\bar{\omega}+\sqrt{\Omega_{-1}^2+\Delta_{-1}^2}$), while deep in the dynamical ferromagnetic phase, the Bloch vector only cycles around the south pole ($n_\mathrm{neff}^* \approx n_\mathrm{neff}^0$). By numerically varying $\beta$ and $V$, we compute the number of roots to produce the phase diagram featured in the main text. If $\beta<0.32$, $f(n_\mathrm{eff})$ can only possess two roots, resulting in a smooth crossover rather than a phase transition.
% The dynamical phase transition hinges on the number of roots for $f(n_\mathrm{eff})$, as depicted in Fig.~\ref{fig:root}. Two key observations can be made regarding function $f(n_\mathrm{eff})$: 1. for the initial state, $f(\bar{\omega}-\Delta_{-1})=f(J_{0,0})=0$, meaning $n_\mathrm{eff}=\bar{\omega}-\Delta_{-1}$ is a solution for $f(n_\mathrm{eff})=0$; 2. To ensure a physical solution for Eq.~\eqref{eq:eomo}, $f(n_\mathrm{eff})$ must be non-positive. As illustrated in Fig.~\ref{fig:root}, $f(n_\mathrm{eff})$ can yield either two or four solutions within the region $n_\mathrm{eff}\in[\bar{\omega}-\sqrt{\Omega^2_{-1}-\Delta^2_{-1}},\bar{\omega}+\sqrt{\Omega^2_{-1}+\Delta^2_{-1}}]$. Physical dynamics occur between the first ($n_\mathrm{eff}=\bar{\omega}-\Delta_{-1}$) and second solutions,

The LMG model~\cite{dpt1,dpt2} supports a  dynamical phase transition with Hamiltonian:
\begin{equation}
    H_\mathrm{LMG}=\chi \hat{S}_z^2  + \Omega \hat{S}_x - \delta \hat{S}_z.
\end{equation}
To gain more insights on how our model related to the LMG model, we can expand Eq. \eqref{eq:h2} to first order and second order in ($2\Omega_{-1} \hat{S}_x + 2\Delta_{-1} \hat{S}_z$). To the first order:
\begin{equation}
    \hat{H}_{\mathrm{eff}}=-\omega_B \hat{S}_z + V (2\Omega_{-1} \hat{S}_x + 2\Delta_{-1} \hat{S}_z) - \frac{V\beta}{N} (2\Omega_{-1} \hat{S}_x + 2\Delta_{-1} \hat{S}_z)^2,
\end{equation}

If we  perform a  rotation along y-axis with angle $\theta=\arctan\Omega_{-1}/\Delta_{-1}$, then  the  Hamiltonian become:
\begin{equation}
\begin{aligned}
   \hat{H}_{\mathrm{eff}}& =-\omega_B \frac{\Delta_{-1}\hat{S}_z - \Omega_{-1} \hat{S}_x}{\sqrt{\Omega^2_{-1}+\Delta^2_{-1}}} +  2 V\sqrt{\Omega^2_{-1}+\Delta^2_{-1}} \hat{S}_z - 4 \frac{V\beta}{N} (\Omega^2_{-1}+\Delta^2_{-1}) \hat{S}_z^2 \\
    &=- 4 \frac{V\beta}{N}  (\Omega^2_{-1}+\Delta^2_{-1}) \hat{S}_z^2 + \frac{\omega_B\Omega_{-1}}{\sqrt{\Omega^2_{-1}+\Delta^2_{-1}}} \hat{S}_x - \Big( \frac{\omega_B\Delta_{-1}}{\sqrt{\Omega^2_{-1}+\Delta^2_{-1}}} - 2 V_1 \sqrt{\Omega^2_{-1}+\Delta^2_{-1}}\Big ) \hat{S}_z,
\end{aligned}
\end{equation}
which takes the form of the LMG model and gives $\tilde{\chi}$, $\tilde{\Omega}$, $\tilde{\delta}$ defined in the main text.
% To the second order:
% \begin{equation}
%     H_{\mathrm{eff}}^{(2)}=-\omega_B \hat{S}_z +  V_1 (2\Omega_{-1} \hat{S}_x + 2\Delta_{-1} \hat{S}_z) + V_2 (2\Omega_{-1} \hat{S}_x + 2\Delta_{-1} \hat{S}_z)^2 + V_3 (2\Omega_{-1} \hat{S}_x + 2\Delta_{-1} \hat{S}_z)^3,
% \end{equation}
% with:
% \begin{equation}
%     V_1 = \frac{1-\beta\bar{\omega}}{(1+\beta \bar{\omega})^3} V,\quad V_2= \frac{-2\beta+\beta^2\bar{\omega}}{(1+\beta \bar{\omega})^4}V, \quad V_3 = \frac{3\beta^2-\beta^3\bar{\omega}}{(1+\beta \bar{\omega})^5} V,
% \end{equation}
% Following the general setup above for the full model, we can find the energy conservation for the first order expansion to be
% \begin{equation}
%     -\omega_B s_z + 2 (V_1 o + V_2 o^2) = \omega_B + 2 (- V_1 \Delta_{-1}  + V_2 \Delta_{-1}^2),
% \end{equation}
% and for the second order:
% \begin{equation}
%      -\omega_B s_z + 2 (V_1 o + V_2 o^2 + V_3 o^3) = \omega_B + 2 (- V_1 \Delta_{-1}  + V_2 \Delta_{-1}^2 - V_3 \Delta_{-1}^3).
% \end{equation}
% Similar as the full model, one can calculate the phase boundary for these approximated spin models.

\section{SCHWINGER BOSONS AND UNDEPLETED PUMP APPROXIMATION}

In this part, we start with the effective Hamiltonian [Eq. \eqref{eq:heff}], but  consider the shallow lattice region around $6E_R$.  The associated  $J_{m,n}$  coumplings are plotted in the right panel of Fig. \ref{fig:Jmn}. Instead of being localized in a single lattice site, the WS states can extend over a few adjacent lattice sites in a shallow lattice.  This can lead to significant suppression of differential AC Stark shifts (homogeneous $J_{n,n}$) at the so-called magic lattice depth ($V_0=6E_R$ in our case). Note that the energy difference between nearest-neighbour WS states is rough $\omega_B$, which allows us to study Bloch oscillations under cavity-mediated interaction. We  consider the WS states with index $m,n\in\{-1,0,1\}$ and use undepleted pump approximation (UPA) $\hat{c}_0\approx\sqrt{N}$ which is valid at  short times when  starting  from $\left|\phi_0\right\rangle$,
\begin{align}
    \hat{N}_{\mathrm{eff}} &\approx 2\Delta_1 \hat{c}^{\dagger}_1 \hat{c}_1 + 2\Delta_{-1}  \hat{c}^{\dagger}_{-1} \hat{c}_{-1} + \sqrt{N} [\Omega_1 (\hat{c}^{\dagger}_1 + \hat{c}_1) - \Omega_{-1} (\hat{c}^{\dagger}_{-1} + \hat{c}_{-1}) ] + NJ_{0,0} \\
    &\equiv \hat{O} + NJ_{0,0}.
\end{align}
Since  $\hat{O}$ is small under UPA  that assumes the $\hat{c}_{\pm 1}$ modes remain almost unoccupied, we can expand the effective Hamiltonian [Eq. \eqref{eq:heff}] up to second order in $\hat{O}$, and  ignore the higher-order terms. The term $\hat{O}^2$ we can be approximated to be:
\begin{equation}
    \hat{O}^2 \approx N [\Omega_1 (\hat{c}^{\dagger}_1 + \hat{c}_1) - \Omega_{-1} (\hat{c}^{\dagger}_{-1} + \hat{c}_{-1}) ]^2,
\end{equation}
and then the effective Hamiltonian becomes:
\begin{align}
    \hat{H}_{\mathrm{eff}} / \hbar &= \omega_B (\hat{c}^{\dagger}_1 \hat{c}_1 - \hat{c}^{\dagger}_{-1} \hat{c}_{-1}) - \frac{VN/\beta}{1 + \beta J_{0,0} + \beta \hat{O} /N} \\
    &\approx \omega_B (\hat{c}^{\dagger}_1 \hat{c}_1 - \hat{c}^{\dagger}_{-1} \hat{c}_{-1}) + V_1 \hat{O} + V_2 \hat{O}^2 \\
    &=\omega_B (\hat{c}^{\dagger}_1 \hat{c}_1 - \hat{c}^{\dagger}_{-1} \hat{c}_{-1}) + V_1 \sqrt{N} [\Omega_1 (\hat{c}^{\dagger}_1 + \hat{c}_1) - \Omega_{-1} (\hat{c}^{\dagger}_{-1} + \hat{c}_{-1}) ] \\
    &+ V_1 (\Delta_1\hat{c}^{\dagger}_1 \hat{c}_1 + \Delta_{-1}\hat{c}^{\dagger}_{-1} \hat{c}_{-1}) + V_2 N [\Omega_1 (\hat{c}^{\dagger}_1 + \hat{c}_1) - \Omega_{-1} (\hat{c}^{\dagger}_{-1} + \hat{c}_{-1}) ]^2 \label{eq:hshort},
\end{align}
here 
\begin{equation}
    V_1 = \frac{2V}{(1 + \beta J_{0,0})^2}, \quad  V_2 = -\frac{V\beta/N}{(1 + \beta J_{0,0})^3}.
\end{equation}
 Moreover, we can absorb the linear term generated by single-particle tunneling via a displacement of a coherent state, $\hat{c}_{\pm 1}=\alpha_{\pm 1}+\hat{c}'_{\pm 1}$ to obtain,
 \begin{equation}
     \hat{H}_{\mathrm{eff}}/\hbar \approx  \omega_B (\hat{c}'^\dagger_{1} \hat{c}'_1  - \hat{c}'^\dagger_{-1} \hat{c}'_{-1}) + V_1 \Delta (\hat{c}'^\dagger_{1} \hat{c}'_1 + \hat{c}'^\dagger_{-1} \hat{c}'_{-1}) 
    + V_2 N \Omega^2 (\hat{c}'^\dagger_{1} + \hat{c}'_1 - \hat{c}'^\dagger_{-1} - \hat{c}'_{-1})^2
 \end{equation}
here we have made the  approximation $\Omega_1 \equiv \Omega \approx -\Omega_{-1}$ as well as $\Delta_1 \equiv \Delta = \Delta_{-1}$. The displacements then become,
\begin{equation}
    \alpha_1 = \frac{V_1 \sqrt{N} \Omega (V_1 \Delta - \omega_B)}{(V_1 \Delta)^2 - \omega_B^2 - 8 \Omega^2 V_2 N  \omega_B },\quad \alpha_{-1} = \frac{V_1 \sqrt{N} \Omega (V_1 \Delta + \omega_B)}{(V_1 \Delta)^2 - \omega_B^2 - 8 \Omega^2 V_2 N  \omega_B }
\end{equation}

The short-time dynamics [Eq. \eqref{eq:hshort}] can be calculated analytically for the quadratic Hamiltonian in terms of  $\hat{c}'_{\pm1},\hat{c}'^{\dagger}_{\pm1}$ with $\hat{c}'_{\pm1} = i[ \hat{H}_{\mathrm{eff}}/\hbar, \hat{c}_{\pm1}]$ and $\hat{c}'^{\dagger}_{\pm1} = i[ \hat{H}_{\mathrm{eff}}/\hbar, \hat{c}'^{\dagger}_{\pm1}]$. In this limit the dynamics   is given by  the equation:
\begin{equation}
i\dot{\left(\begin{array}{c}
\hat{c}'_{1}\\
\hat{c}'_{-1}\\
\hat{c}'^{\dagger}_{1}\\
\hat{c}'^{\dagger}_{-1}
\end{array}\right)}=\mathcal{H}_{\mathrm{BdG}}\left(\begin{array}{c}
\hat{c}'_{1}\\
\hat{c}'_{-1}\\
\hat{c}'^{\dagger}_{1}\\
\hat{c}'^{\dagger}_{-1}
\end{array}\right),\label{eq:upa}
\end{equation}
with the coupling matrix $S$:
\begin{equation}
   \mathcal{H}_{\mathrm{BdG}} =\left(\begin{array}{cccc}
\omega_{B}+V_{1}\Delta+2V_{2}\Omega^{2} & 2V_{2}\Omega^{2} & -2V_{2}\Omega^2 & -2V_{2}\Omega^2\\
-2V_{2}\Omega^2 & -2V_{2}\Omega^2 & -\omega_{B}+V_{1}\Delta+2V_{2}\Omega^{2} & 2V_{2}\Omega^{2}\\
-2V_{2}\Omega^{2} & -\omega_{B}-V_{1}\Delta-2V_{2}\Omega^{2} & 2V_{2}\Omega^2 & 2V_{2}\Omega^2\\
2V_{2}\Omega^2 & 2V_{2}\Omega^2 & -2V_{2}\Omega^{2} & \omega_{B}-V_{1}\Delta-2V_{2}\Omega^{2}
\end{array}\right).
\end{equation}
The matrix $\mathcal{H}_{\mathrm{BdG}}$ can have either real or complex eigenvalues, which leads to distinct dynamical behaviors as shown in the main text. When all the eigenvalues are real, the populations $\rho_{\pm1}$,  with  $\rho_{n}=\langle \hat{c}_n^{\dag}\hat{c}_n\rangle$,
feature stable small amplitude oscillations; on the other hand when  all the eigenvalues are complex, then $\rho_{\pm1}$ feature  an exponential growth associated with the
correlated pair production of atoms at WS centered at $n=\pm 1$,   which leads to the amplification of the Bloch oscillation signal until the  UPA breaks down. 
% The four eigenvalues are:
% \begin{equation}
%     \eta = \pm \sqrt{(V_1 \Delta)^2 + 2 V_1 \Delta V_2 \Omega^2 + \omega_B^2 \pm \sqrt{V_1 \Delta [V_1 \Delta (V_2 \Omega^2)^2 + V_1 \Delta \omega_B^2 + 2V_2 \Omega^2 \omega_B^2]} }.
% \end{equation}
% Actually it will be hard to analyse due to $V_{1,2}$ depend on $\beta$ and lattice depth $V_0$ (by $J_{0,0}$), at the same time $\Delta$ and $\Omega$ also depend on $V_0$. However we know that when increasing $V_0$ across magic lattice depth $6E_R$, $\Delta$ change from negative to positive value and then the square root inside will be negetive due to $V_2<0$.
   
\section{Experimental considerations}
\subsection{Single-particle Bloch oscillations}
In this section, we discuss the protocols to observe single-particle Bloch oscillations in the experiment. The main idea is to prepare a superposition of different WS states which accumulate different phases under $\hat{H}_0$. In the main text, we discussed the quench scheme where  the initial localized WS state $\phi_0$ becomes a superposition of delocalized WS states.  An alternative way to probe Bloch oscillations is to  amplitude modulate  the lattice depth as:
\begin{equation}
\begin{aligned}
     H_1(t) /\hbar &= V_{1}\sin^{2}k_{l}z\cos(\omega t+\phi) \\
     &=\sum_{m,n=-\infty}^{\infty}t_{m}\cos(\omega t+\phi)(\hat{c}_{m+n}^{\dagger} \hat{c}_n + \hat{c}_{n}^{\dagger} \hat{c}_{m+n}),
\end{aligned}
\end{equation}
which has been demonstrated in~\cite{bo9}. Here we define the tunnelling rate between $\phi_{m+n}$ and $\phi_n$ as:
\begin{equation}
    t_{m}=V_{1}\int dz\sin^{2}(k_{l}z)\phi_{m+n}(z)\phi_{n}(z),
\end{equation}.

Moreover, we can choose $\omega\approx m \omega_B, m \in \mathbb{Z}$ to drive the  $m$th sideband (between $\phi_{m+n}$ and $\phi_{n}$) and ignore the fast rotating terms:
\begin{equation}
\begin{aligned}
         H_1(t)/\hbar &=\sum_{m,n=-\infty}^{\infty}t_{m}\cos(\omega t+\phi)(e^{im\omega_{B}t}\hat{c}_{m+n}^{\dagger} \hat{c}_n +e^{-im\omega_{B}t}\hat{c}_{n}^{\dagger} \hat{c}_{m+n}) \\
         &\approx \sum_{n=-\infty}^{\infty}\frac{t_{m}}{2}(e^{-i\phi}\hat{c}_{m+n}^{\dagger} \hat{c}_n+e^{i\phi}\hat{c}_{n}^{\dagger} \hat{c}_{m+n}).
\end{aligned}
\end{equation}

\begin{figure}
	\includegraphics[width=0.4\columnwidth]{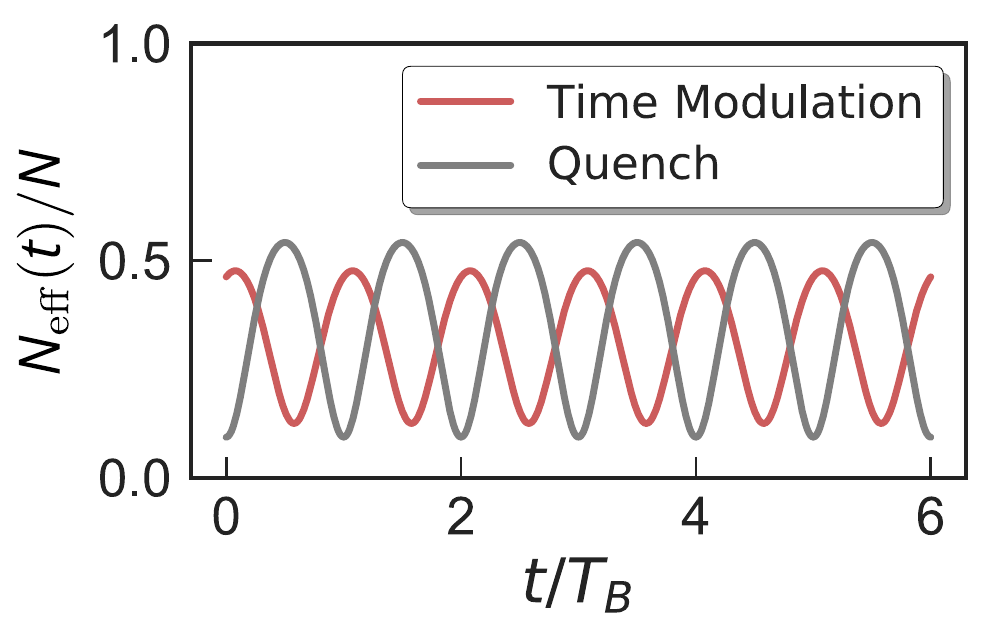}
	\caption{\label{fig:single} Single particle Bloch oscillation with amplitude modulation scheme (pink curve, $V_0=8E_R$ and $V_1=0.4E_R$) and Quench scheme (grey curve, quench from $V_0=15E_R$ to $V_0=8E_R$ as main text).
	}
\end{figure} 

As a result, starting from $\phi_0$ and performing  the amplitude modulation for time $\tau$, we obtain the initial state to be a superposition of WS states $\{\phi_{n\times m}\}$. In Fig.~\ref{fig:single} (pink curve), we simulate the case with lattice depth $V_0=8E_R$ and modulation strength $V_1=0.4E_R$, also the first sideband transition ($\omega=\omega_B$). Different from the quench scheme (grey curve), after the modulation the single particle wavefunction can have a  non-zero coupling to the cavity field ($N_{\mathrm{eff}}/N\neq0$).

In the experiment, there may be higher bands populated  in the quench protocol we discussed in the main text. In other words, the WS basis describing the ground band for the shallow lattice (after quench) is not necessarily complete to describe the initial localized state. Higher bands population will inevitably introduce other frequency components to $N_{\mathrm{eff}}(t)$  disrupting the BO signal. However, in the simulations we performed for the main text (quench from $V_0=15E_R$ to $V_0=8E_R$), $98\%$ atoms remained in the ground band and the higher band population can be ignored. Similarly, in the amplitude modulation schemes, we also choose $V_1$ to be much smaller than the band gap to avoid higher bands population.

\subsection{Experimental parameters}
Here we discuss the parameters for the  specific case of $^{87}\mathrm{Rb}$ with incommensurate lattice wavelength ($\lambda_l=532$ nm, $\omega_B=2\pi\times557$ Hz) and cavity wavelength ($D_2$ transition with $\lambda_c=780$ nm). 
We are interested in the parameter regime with $V\sim\omega_B$ and $\beta\sim O(1)$, where the  dynamics is mostly unitary and the dissipative processes can be ignored as we explain below. Another requirement is  that the band gap ($27\omega_B$ for $\lambda_l=532$ nm and $V_0=6E_R$) should be much larger than $V$ if we want to only work with the ground band WS states.
Moreover, the cavity decay rate $\kappa \sim 2\pi \times 0.1$ MHz, the atom-light coupling strength $\mathcal{G}_0 \sim 2\pi\times 0.3$ MHz, and atomic transition decay rate $\gamma \sim 2\pi\times 10$ MHz give the cavity cooperativity $C=4\mathcal{G}_0^2/\gamma\kappa\sim0.36$, which can be tuned even larger for larger $\mathcal{G}_0$ and smaller $\kappa,\gamma$. The cavity loss generates collective dephasing processes at a rate $V\beta \kappa/\Delta_c$, while spontaneous emission generates off-resonant photon scattering
processes at a rate $V\gamma/\Delta_0$ as mentioned in the main text.
% To make it more intuitive, we can reexpress $\beta=N\mathcal{G}_0^2/\Delta_c\Delta_0=(NC/4)(\gamma/\Delta_0)(\kappa/\Delta_c)$. 
Under $\kappa/\Delta_c\sim0.05$ and $\gamma/\Delta_0\sim0.01$, one obtains negligible dissipation within the experimentally relevant time scales and $\beta\sim O(1)$. For the maximum AC Stark shift, we first find that $\mathcal{G}_0^2/\Delta_0\sim2\pi\times100$ Hz with the parameters listed above, then $|\eta_p|^2/\Delta_c^2$ can be tuned between $1$ to $10$ for $V\sim\omega_B$.

Our proposal works with a single internal level in the ground state manifold for atoms hopping between motional states (WS states here). Since interactions are mediated by photons, quantum statistics are not important in our scheme. As a result, even though above we considered the case of Rb, our model can be realized with other species of alkali atoms ($D_2$ transition) and alkaline earth atoms (${}^1S_0\rightarrow {}^3P_1$ transition) i.e. $^{87}\mathrm{Sr}$ (boson), $^{88}\mathrm{Sr}$ (fermion), $^{171}\mathrm{Yb}$ (fermion) with appropriate choices of lattice wavelength and magic lattice depth summarized in table~\ref{table:1}.  Note that both $^{88}\mathrm{Sr}$,  $^{171}\mathrm{Yb}$  have very small scattering lengths in the ground states. 

The single particle Bloch oscillations and dynamical phase transition in the deep lattice doesn't set too much limit on the choice of $\lambda_l$ and $\lambda_c$. We only want the near-neighbour coupling coefficient $J_{m,m+1}$ to be larger, while the overlaps between $\phi_m(z)$, $\phi_{m+1}(z)$ and $\sin^2(k_c z)$ become tiny when $\lambda_l\approx\lambda_c$, so we want to choose different $\lambda_l$ and $\lambda_c$. While for the amplification of BOs in the shallow lattice region, we need to perform the experiment around magic lattice depth thus too shallow magic depth (such as $^{171}\mathrm{Yb}$) isn't favorable.

%In the experiment, one can observe dynamical phase transition under deep lattice and amplification of BO under shallow lattice (around the particular magic lattice depth) with different atomic species.

\begin{table}
\begin{tabular}{|c|c|c|c|}
\hline 
Atomic species & $\lambda_{l}$ (nm) & $\lambda_{c}$ (nm) & Magic lattice depth ($E_{R}$)\tabularnewline
\hline 
$^{87}\mathrm{Rb}$ (boson) & 532 & 780 & 6$E_R$ \tabularnewline
\hline 
$^{87}\mathrm{Sr}$ (fermion) & 532 & 689 & $5E_R$ \tabularnewline
\hline 
$^{88}\mathrm{Sr}$ (boson) & 532 & 689 & $5E_R$ \tabularnewline
\hline 
$^{171}\mathrm{Yb}$ (fermion) & 413 & 556 & $3.2E_R$ \tabularnewline
\hline 
\end{tabular}
\caption{Summarized lattice, cavity wavelength and magic lattice depth for different atomic species.}
\label{table:1}
\end{table}

\subsection{Radial mode thermal distribution}
In this section, we discuss the effect of the Radial thermal motion  following  Ref.~\cite{dpt3}. The Gaussian geometry of the laser beams in experiments inevitably couples the vertical and radial wave functions. The Gaussian profile of the lattice and cavity beams causes atoms in different radial modes to have different tunneling rate, resulting in a slightly different overlap integral $J_{m,n}$ for atoms in different radial modes. This  effect can also be understood as fluctuations of the lattice potential $V_0$ due to radial thermal excitation.

\begin{figure}
	\centering
	\includegraphics[width=0.8\columnwidth]{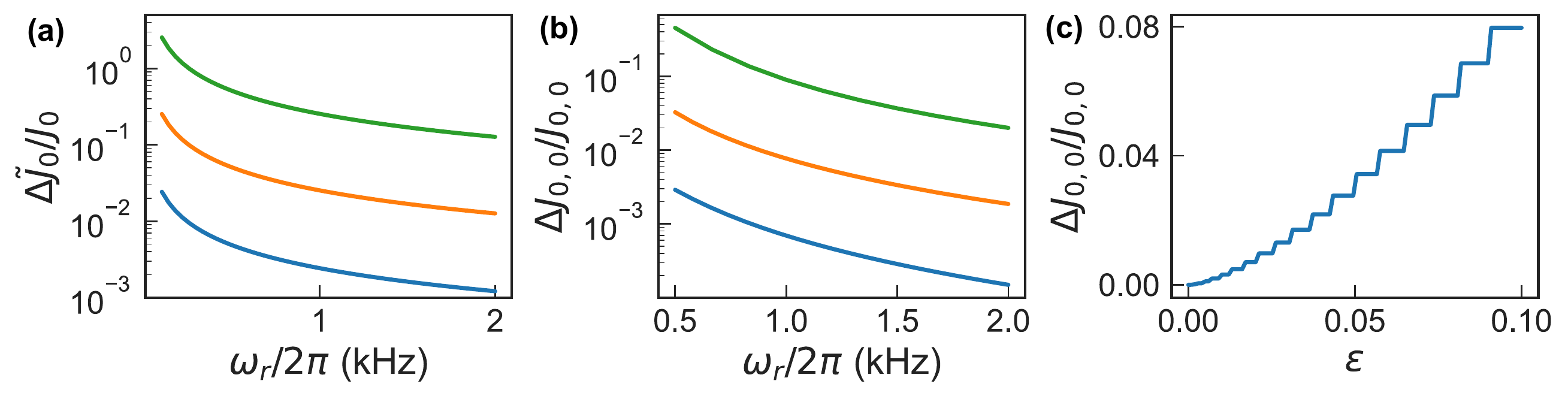}
	\caption{\label{fig:tm} The standard deviations of (a) the ground band tunneling rates, (b) the coupling coefficient $J_{0,0}$ as a function of radial trapping $\omega_r$ with fixed $T=0.1\ \mu$K (blue curve), $T=1\ \mu$K (orange curve) as well as $T=10\ \mu$K (green curve). We use the beam width $w_l=50 \mu$m and lattice potential $V_0=6E_R$ in the calculation. (c) The standard deviations of $J_{0,0}$ as a function of loading error rate $\epsilon$.
 	}
\end{figure} 

First, we focus on the Gaussian beam profile of a 1D lattice, which leads to the following trapping potential:
\begin{equation}
    V_0(r,z)=V_0 \sin ^2\left(k_l z\right) \exp \left(-2 r^2 / w_l^2\right),
\end{equation}
where  $w_l$ is the beam width. In the presence of  additional radial trapping potential $V_r(r,z)=M\omega^2_r r^2 /2$~\cite{cox2016spatially} we can expand the total trapping potential $V(r,z)=V_0(r,z)+V_r(r,z)$ to second order of $r$ and obtain:
\begin{equation}
    V(r,z)\approx V_0 \sin^2(k_l z) + \frac{1}{2} M \omega_r^2 r^2  - \frac{\omega^2_{r0}}{\omega_r^2} \frac{1}{2} M \omega_r^2 r^2\sin^2(k_l z),
\end{equation}
here $\omega_{r0}=\sqrt{4V_0/M w_l^2}$. 

The first term  describes the lattice potential along the axial direction with the characteristic Bloch functions as eigenstates. The second term describes the radial harmonic trapping with eigenstate $\phi_{n_x,n_y}(r)=\phi_{n_x}(x)\phi_{n_y}(y)$ and eigenenergies $E_{n_x,n_y}=\hbar\omega_r(n_x+n_y+1/2)$. The third term describes the coupling between axial and radial degrees of freedom.  The correction of $J_0$ is given by~\cite{dpt3}:
\begin{equation}
    \tilde{J}_0\left(n_x, n_y\right)=J_0+\frac{1}{8} \frac{\omega_{r 0}^2}{\omega_r^2} E_{n_x, n_y}[\frac{\partial}{\partial v_0} f(\tilde{q}=0, v_0 / 4)-\frac{\partial}{\partial v_0} f(\tilde{q}=\pm 1, v_0 / 4)].
\end{equation}
Here the function $f$ is the characteristic Mathieu value of type A for $q\in (-\hbar k_l,\hbar k_l)$, and the characteristic Mathieu value of type B for $q=\pm k_l$. We define $\tilde{q}=q/\hbar k_l$ and $v_0=V_0/E_R$. One can take such $\tilde{J}_0$ into Eq.~\eqref{eq:ws} to calculate $\phi_n(z)$, which causes inhomogeneity for the coupling matrix $J_{m,n}$. We can estimate the contribution from different radial eigenmode with Boltzmann distribution $p_{n_x,n_y}=\exp[-(n_x+n_y)\omega_r \hbar / k_B T] / Z$, in which the partition function $Z\approx (k_B T/\hbar \omega_r)^2$. Then we can calculate the variance of the $\tilde{J}_0$ as:
\begin{equation}
    \begin{aligned}
             \Delta \tilde{J}_0^2 &= \sum_{n_x,n_y} p_{n_x,n_y}  [\tilde{J}_0\left(n_x, n_y\right) - J_0]^2 \\
             &= \{\frac{\hbar}{8}\frac{\omega_{r0}^2}{\omega_r} [\frac{\partial}{\partial v_0} f(\tilde{q}=0, v_0 / 4)-\frac{\partial}{\partial v_0} f(\tilde{q}=\pm 1, v_0 / 4)]\}^2 \frac{2(e^{\hbar\omega_r/k_BT} +2)}{(e^{\hbar\omega_r/k_BT}  -1)^2}.
    \end{aligned}
\end{equation}

In Fig.~\ref{fig:tm}(a), we plot the standard deviation of the tunneling rate $J_0$ as a function of $\omega_r$ and different temperature $T$. In Fig.~\ref{fig:tm}(b), we plot the standard deviation of the coupling coefficient $J_{0,0}$ due to the correction of the tunneling rate. Similar behavior for other coupling coefficients $J_{m,n}$. As a result, one can suppress the effect of the radial modes  occupation by increasing the total radial trapping frequency $\omega_r$ or lowering the temperature. The standard deviation $\Delta J_{0,0 }\approx 0.01 J_{0,0}$ up to
temperature $T\sim 1\ \mu$K as well as $\omega_r=2\pi\times1$ kHz, thus the radial thermal noise only has a tiny effect on many-body dynamics we predict.

\subsection{Atoms loading}
In this section, we discuss the real process of atoms loading in the experiment. In the main text, we mention first loading atoms at position $k_c z /\pi =r, r\in\mathbb{Z}$ which atoms-cavity coupling becomes perfect zero. However one can only set a threshold for the atom-cavity coupling during the loading process i.e. load all the atoms with $\sin^2 k_c z < \epsilon$ in the real experiment. Such loading error makes $J_{m,n}$ deviate from expected values, which brings additional inhomogeneity. In Fig.~\ref{fig:tm}(c), we plot the standard deviation of the coupling coefficient $J_{0,0}$ as a function of error $\epsilon$.  We consider the total lattice length to be 1 mm and assume atoms load into all the sites $n$ which satisfy $\sin^2(k_c n a_l)<\epsilon$ uniformly. These imperfect sites cause tiny inhomogeneity in the coupling coefficient $J_{m,n}$ up to $\epsilon\sim5\%$.

\bibliography{apssamp}